\documentclass{aastex631}

\usepackage{multirow}

\definecolor{Red}{rgb}{0.743,0,0}
\definecolor{Blue}{rgb}{0.25,.41,.88}
\definecolor{Green}{rgb}{0,0.5,0}

\shorttitle{On the Ordering of Exoplanet Systems}
\shortauthors{Lozovsky \& Perets}

\graphicspath{{./}{figures/}}
\begin{document}
%On the ordering and architecture of multiple-planet exoplanet systems
\title{On the Ordering of Exoplanet Systems}

\author[0000-0001-5451-3221]{Michael Lozovsky}
\affiliation{Astrophysics Research Center of the Open University (ARCO),\\
Department of Natural Sciences,\\
The Open University of Israel \\
4353701, Raanana, Israel}

\author[0000-0002-5004-199X]{Hagai B. Perets}
\affiliation{Department of Physics, Technion - Israel Institute of Technology, \\
Haifa, 3200002, \\
Israel
}
\affiliation{Astrophysics Research Center of the Open University (ARCO),\\
Department of Natural Sciences,\\
The Open University of Israel \\
4353701, Raanana, Israel}

\begin{abstract}
We present a comprehensive analysis of planetary radii ordering within multi-planet systems, namely their ordinal position with respect to their size in a given system, utilizing data from the NASA Exoplanet Archive. In addition, we consider not only the ordinal positions but also the specific period ratios and radius ratios of planetary pairs in multi-planet systems. We explore various dependencies on stellar host type and metallicity, as well as planetary types, and explore the differences between planetary systems with different planet multiplicities and different planetary pairs in the same system. Focusing on \textit{Kepler} systems with two to four planets, we account for observational biases and uncover a robust trend of smaller inner planets. This trend is particularly pronounced in inner pairs of three-planet systems and exhibits variations in stellar metallicity and planet multiplicity. Notably, we find that the distribution of inner-to-outer planet radii ratios depends on the system's metallicity, suggesting a link between initial conditions and the resulting system architecture. Interestingly, planet pairs in resonance do not exhibit significantly different size ratios compared to non-resonant pairs, challenging current theoretical expectations, again,  possibly suggesting that initially resonant systems could have been later destabilized.
Our findings align with planet formation and migration models where larger planets form farther out and migrate inward. Importantly, we emphasize the significance of planet ordering as a novel and crucial observable for constraining planet formation and evolution models. The observed patterns offer unique insights into the complex interplay of formation, migration, and dynamical interactions shaping planetary systems.

\end{abstract}


\section{Introduction} \label{sec:intro}

The last few decades of space exploration led to the discovery of thousands of exoplanets, allowing for the first time the statistical characterization of exoplanetary systems.  The wide range of observational tools and methods allow for the characterization of different properties, including planetary sizes, masses and orbits \citep[e.g.][]{Lozovsky2017,Mazevet2023,Helled2016,Venturini2024,Dorn2017,Lammers2023}, and to a lesser extent even planetary atmospheres\citep[e.g.][]{Konopacky2013,Bourrier2020,Calvin2021,Shi_Yuan2021,Almenara2022,Tinetti2018} and spins \citep[e.g.][]{Triaud2010,Snellen2014}. 
Currently, most exoplanetary systems were discovered through the transit method, and in particular by space missions such as \textit{Kepler} (and \textit{K2}), \textit{CHEOPS} and \textit{TESS} \citep[e.g.][]{Van_Grootel2021,Thuillier2022}.

A wide diversity in the properties and the system architectures of identified systems has been found \citep{Dorn2017,Agol2021,Mishra2023,Howe2025}, suggesting a wide range of planetary structures and compositions, as well as complex planet formation and dynamical evolution processes taking place during the assemble of planetary systems.

Multi-planet systems exhibit various hierarchical structures, that do not resemble our own Solar System. Patterns in these multi-planet architectures may provide important hints about the formation and evolution of planets surrounding different stars \citep[e.g.][]{Weiss2020,Zhu2019}. The architectures we observe are the product not only of formation but also subsequent dynamical evolution. One cannot reconstruct the initial conditions and formation history directly from planetary size ordering, as dynamical interactions change the initially formed systems, both during the formation of the planets and afterward \citep[e.g.][]{Deck2012,Kipping2017}. However, planet size ordering may hold major and unique clues about the various planet formation and dynamical evolution of these systems, which differ from the statistical information given by the overall statistics on individual planets. 

Various studies applied statistical analysis to exoplanetary data and found interesting trends in the data, that can be linked to formation mechanisms of planetary bodies and planetary systems and therefore give few hints on their internal structure \citep[e.g.][and references within]{Ciardi2013,Weiss2018,Otegi2020,Burke2015,Helled2016,Kipping2017,Sandford2021, Mishra2023, Venturini2024,schulze2024}. However, it is important to keep in mind that the discovered planetary properties might suffer from instrumental and human selection biases \citep{ANANYEVA2020,Pascucci2018,Gaudi2005,Petigura2013,Thomas2025}. Furthermore, the data samples are from non-homogeneous sources, with different biases. Most exoplanets discovered today originated from the \textit{Kepler} mission and were discovered via the transit method. Therefore, the \textit{Kepler} sample might be considered homogeneous regarding well-known selection biases \citep{Kipping2016}, which can be partially accounted for.

%Thus,\textit{ Kepler} sample might be considered  homogeneous in a sense of well-known selections biases \citep{Kipping2016}, which could be accounted for, at least partially. 

There is an ongoing debate whether the trends discovered in \textit{Kepler}'s multi-planet systems represent a physical tendency or a selection bias. \citet{Weiss2018} suggested that planets in a \textit{Kepler} multi-planet system are similar in size and regularly spaced, in the so-called "peas in the pod" hypothesis. This finding was challenged, revised \citep{Chevance2021,Otegi2021,Zhu2020,Mishra2021}, justified and reconfirmed \citep{Murchikova2020,Weiss2020,Fred2020,Lammers2023,Otegi2022,Mamonova2024,Goyal2024,Chance2024,Goldberg2022,Thomas2025} by different groups. Previously, a different approach to the same question was presented by \citet{Millholland2017}, who studied \textit{Kepler}'s planet masses found via transit time variations (\textit{TTV}) from \citep{Hadden2017}. \citet{Millholland2017} found that planets orbiting the same star also tend to have similar masses, strengthening the later claim of \citet{Weiss2018}.

The currently observed hierarchical structure of a system likely depends on the formation environment of the system. Planets are born in protoplanetary discs \citep{Pollack1996,Safranov1969,Raymond2022}, and some properties of protoplanetary disks could still be imprinted on the exoplanet population \citep{Mulders2021}. The mass distribution in a protoplanetary disk  \citep[e.g.][]{Manara2018,Tripathi2017} is possibly reflected in the exoplanet population, with the larger planets preferentially located farther out than smaller ones. However, we should keep in mind that one cannot infer the exact properties of  protoplanetary disk directly from planetary composition \citep[e.g.][]{Lozovsky2022,Lozovsky2023, Moriarty2014,Thiabaud2015}

Statistical studies indicate that the sizes of planets in exoplanetary systems are not randomly distributed, but present some correlations and trends \citep[e.g.][]{Mazeh2016,Kipping2017,Cabot2022,Biazzo2022}. For instance, \citet{Helled2016} found a correlation between planetary radius and orbital period; \citet{Mulders2021} found a correlation between the masses of planetary systems and their respective orbital periods. Some studies found dependence between stellar metallicity and planetary properties \citep[e.g.][and references therein]{owen2018metallicity,Thorngren2016}.
Surveys have shown that exoplanets that are smaller than Neptune are common at orbital periods less than a year around Sun-like stars \citep{Petigura2013,Fressin2013}.
It was found by \citet{Lissauer2011} that the size \textit{Kepler}'s planet is correlated with the size of its detected neighbors. \citet{Lissauer2011} found an anti-correlation between the mean density of the planets and the orbital period meaning that the lower-density planets are located farther out than the smaller and denser planets. Later-on \citet{Ciardi2013} showed that for planet pairs the larger one most often will have a longer period, for pairs in which one or both planets are approximately Neptune-sized or larger. 

%Most of these studies have focused on the statistics of the planets in systems, but did not consider relative data of planets in the same system, and in particular the size/mass ordering of planets in the same system, which is the focus on the current study.
The majority of previous studies have primarily examined the statistical characteristics of planets within systems, without specifically analyzing the relative data of planets within a given system; nevertheless, some studies explored multi-planet systems characteristics \citep{Ciardi2013,Kipping2017,He+20,Mishra2023,Chance2024}, though mostly focusing on two-planet systems. In particular, the size/mass ordering of planets within a single system, which is the central focus of the current study, has been little explored.
Here we focus on transiting multi-planet systems and investigate the relative sizes of planets in every system in a sample of exoplanets from \textit{NASA} database. 

In addition, it is important to note that the planetary properties might be linked to stellar properties, which by themselves relate to the protoplanetary disk properties, e.g. in terms of metallicity \citep{Muller2022,Mishra2023,Dawson2013} or mass/stellar-type, and thereby also relate to the location in the Galaxy, \citep[e.g.][]{Bashi2022}. So, it is reasonable to assume that planetary statistics and the hierarchical structure of a system might be influenced by the hosts properties. As a part of this study, we divide the planetary sample by the metallicity of the stars and show the differences between the populations, as well as explore the dependence on stellar type. 

In this paper, we first briefly review the physical processes that can affect planetary ordering in section \ref{sec:processes}, we then discuss our data collection and analysis methods in section  \ref{sec:methods}, followed by the presentation of our analysis results (section \ref{sec:results}), and finally the discuss and summary in section \ref{sec:Conclusions}.

%We show that for the inner planets indeed tend to be smaller, but this tendency that exists in two-planet systems is not so clear in higher orders.

%Understanding the relative sizes of the planets within a system can yield new clues on the formation, migration, and evolution of planets within a system, and hence to understand the physical phenomena shaping the planetary systems.

\section{Physical processes sculpturing the architecture and ordering of planetary systems}
\label{sec:processes}

The architecture of a planetary system is sculpted by a combination of early formation processes and subsequent dynamical evolution, including migration \citep[e.g.][]{Ford2014,Emsenhuber2023}, planet-planet interactions\citep[e.g.][]{Mishra2023}, evaporation \citep[e.g.][]{Innes2023,Mordasini2020}, and tidal effects \citep[e.g.][]{Papaloizou2016}. These processes are not mutually exclusive and can act concurrently. We briefly discuss how these mechanisms might influence the observed ordering of planets, emphasizing that they contribute to general trends, while the specific outcome for a given system depends on the interplay of various factors.

\subsection{Initial Formation}

The initial configuration of planetary systems is established during planet formation within protoplanetary disks. Two primary models are considered:

\begin{itemize}
\item \textbf{Gravitational instability (GI) and/or turbulent fragmentation}: This mechanism involves the direct collapse of disk material into giant planets, primarily in the outer regions of massive disks \citep{Boss1997}. While \textit{Kepler} observations favor detecting planets in the inner regions and may be less sensitive to GI-formed planets at large distances, this scenario could contribute to the{{so-called brown dwarf desert \citep[e.g.][]{Unger2023, McCarthy2004}, which is a range of orbits around a star, where brown dwarfs are unlikely to be found as companion objects \citep[e.g.][]{Beauge2007}}}. The GI model is less likely to play a dominant role in the ordering of inner, rocky planets.

\item \textbf{Core accretion (CA)}: Planetesimals accumulate to form rocky cores, which accrete gas if massive enough \citep{Pollack1996}. Larger planets tend to form farther from the star due to increased material availability in the outer disk, especially beyond the snowline \citep{Mordasini2009}. In relatively undisturbed systems, we expect a preference for larger planets at larger distances, consistent with our findings of larger outer members in pairs (even after debiasing). 

Stellar metallicity influences core accretion, with higher metallicity promoting faster core formation and potentially more and/or larger planets \citep{Johnson2010}. Consequently, varying metallicities may lead to different planetary ordering due to differing rates of planet-planet interactions and scattering, as also suggested by metallicity-dependent eccentricity distribution \cite{Dawson2013}. 
\end{itemize}

\subsection{Planetary disk migration}

Planetary disk migration, involving planet-disk interactions, alters orbital distances and shapes the system architecture. Two main types are considered in the literature:

\begin{itemize}
    \item \textbf{Type I Migration:} This affects lower-mass planets embedded in the disk, experiencing torques that cause inward or outward migration depending on the disk's properties \citep{Ward1997}.
    \item \textbf{Type II Migration:} More massive planets, capable of opening gaps in the disk, migrate at a rate governed by the disk's viscous evolution \citep{Kley2012}.{{However, some studies suggest that this evolution is a modified version of Type I migration, which accounts for the reduced gas surface density \citep{Kanagawa2020}}}
\end{itemize}

Larger planets forming farther out and migrating inward can explain their presence in regions where in-situ formation is unlikely. While unlikely to alter the initial ordering directly, migration coupled with observational biases can affect the observed ordering. For example, a migrating large outer planet may become observable by Kepler, while a non-migrating smaller planet remains farther out and is less likely to be detected, biasing the observed ordering towards larger outer planets. 

Differential migration rates can lead to resonant capture or scattering, further influencing the ordering.

\subsection{Migration of Resonant Pairs}

Convergent migration leading to resonant capture and divergent migration avoiding resonance depends on factors including:

\begin{enumerate}
    \item \textbf{Mass Ratio:} Convergent migration is favored in pairs with similar masses, with a larger outer planet, while divergent migration is more likely when the inner planet is much more massive \citep{papaloizou_migration_2005, nelson_migration_2000,Won+24}.
    \item \textbf{Disk Properties:} Density and temperature profiles affect migration direction and rate, with steeper profiles favoring convergence \citep{paardekooper_torque_2010}.
    \item \textbf{Resonance Location:} Closer-in resonances are more prone to convergent migration due to stronger planet-disk interactions \citep{ogihara_resonance_2013}.
    \item \textbf{Planet-Disk Interaction:} More massive planets interact more strongly, potentially leading to faster or runaway migration \citep{kley_interaction_2012}.
\end{enumerate}
These effects could lead to specific planet ordering patterns in resonant pairs. For example, faster migration of an outer massive planet could capture an inner planet into resonance and migrate inward together in resonance, while the reverse ordering might not result in such an outcome. One might therefore generally expect a different ordering of planetary pairs/triples in resonances compared with non-resonant systems, with a possible additional preference for larger outer planets in resonant pairs.

\subsection{Planet-planet scattering}
Planet-planet scattering is a significant process that can alter the architecture of planetary systems. Strong gravitational interactions between planets can lead to planet-planet scattering giving rise to orbital instabilities, and resulting in collisions (both mutual and with the host star), ejections, or significant changes and exchanges in orbital configurations \citep[e.g][and references therein]{Chatterjee2008}. This process tends to:
\begin{itemize}
    \item Preferentially retain more massive planets while ejecting smaller ones \citep{Chatterjee2008}
    \item Result in systems with fewer, more widely spaced planets
    \item Produce planets on eccentric orbits
    \item Excite inclinations; which in the context of transiting planets may lead to non-detection of all the planets in a given system, when their orbits are not sufficiently aligned.
    \item Give rise to more massive planets due to mutual collisional growth.
    \item Exchange between inner and outer planets ordering.
\end{itemize}

Planet-planet scattering is likely to be more frequent in planetary systems with a larger number of planets and/or more massive planets, which give rise to larger perturbations. The planet scattering could significantly change the architecture of both for the above reasons.
Given the different dependence on inclinations, one might observe different planetary ordering in systems observed with radial-velocity (RV) measurements which can detect non-aligned planets. Here we focus only on the Kepler transiting systems, but a similar effort should be made with RV-measured systems. The latter, however, currently contain far smaller statistics.  

Generally, systems hosting fewer planets/less-massive planets may experience more quiet dynamical evolution, compared with more massive ones. If more metal-rich systems give rise to a larger number/mass of planets we might expect such systems to experience more interactions, and be less ordered and have larger planets on average.   

\subsection{Secular planet-planet interactions}
Secular interactions between planets occur on timescales much longer than orbital periods and can significantly influence the long-term evolution of planetary systems. These interactions can lead to various phenomena:

\begin{itemize}
    \item \textbf{Secular resonances:} These occur when the precession frequencies of two planets' orbits match, leading to large-scale oscillations in eccentricities and inclinations \citep{Lithwick2012}.
    
    \item \textbf{von-Ziepel-Kozai-Lidov mechanism:} In hierarchical triple systems, this can cause large oscillations in eccentricity and inclination of the inner orbit \citep[see][for a review]{Naoz2016}.
    
    \item \textbf{Secular chaos:} In systems with three or more planets, an overlap of secular resonances can lead to chaotic evolution of eccentricities and inclinations \citep[e.g][]{Lithwick2011chaos, Ham+17}.
\end{itemize}

These secular effects can significantly alter planetary system architectures over long timescales, leading to the excitation of high eccentricities, which might couple with tidal interactions to give rise to eccentric migration \cite{Wu+03}. 

Secular interactions can also influence the mutual inclinations of planets. Systems that have undergone significant secular evolution may have higher mutual inclinations, potentially leading to non-transiting configurations for some planets. This could introduce biases in the observed architectures of transiting systems.

Furthermore, the strength of secular interactions depends on planetary masses and orbital separations. Systems with more closely spaced and/or more massive planets are more likely to experience strong secular effects, potentially leading to different observed architectures compared to more widely spaced or less massive systems. Systems with significantly misaligned planetary orbits are more sensitive to secular van-Ziepel-Lidov-Kozai processes.

Overall, secular processes are less likely, by themselves, to change planetary ordering, unless they lead to crossing orbits and instabilities, which in tune give rise to strong planet-scattering. Nevertheless, the eccentric migration that might be triggered could change the orbits and, like in the case of disk migration lead to an effective observational bias change in ordering through the coupling to the observational bias of detecting only close-in planets.

\subsection{Tidal Evolution and Atmospheric Loss}
For close-in planets or planets with a close pericenter approach tidal interactions with the host star can cause orbital decay, potentially leading to the destruction or engulfment of inner planets \citep[e.g.][]{Wu+03,Bolmont2012}. This process preferentially affects larger, closer (to the star) planets and can alter the observed size ordering over time. Additionally, atmospheric loss due to stellar irradiation may play a role in altering the size distribution of inner planets over time \citep{Owen2013}, further reinforcing the observed size ordering preference towards smaller planets at inner regions.

\section{Methods}
\label{sec:methods}

In this work, we apply a simple statistical analysis of transiting exoplanetary data, based on \textit{NASA} database \citep{NASA_Exoplanet} to explore the observed trends in multiple-planet systems.

The planetary parameters used in the analysis are the planetary radii and orbital periods; the stellar parameters used are effective temperature and metallicity. First, we build an exoplanetary sample based on the database. Next, we classify each planetary system by its planetary multiplicity. Then, we study the relative properties (radii and period) of planets within each system and compare between different classes of systems we define. Comparing planets within a given system eliminates systematic uncertainties associated with stellar properties.

\subsection{Planetary sample}
\label{subsec:Sample}

\begin{figure}[h]
	\centering

	\includegraphics[width=1\columnwidth, trim={0.5cm 6.1cm 1.3cm 7.0cm},clip]{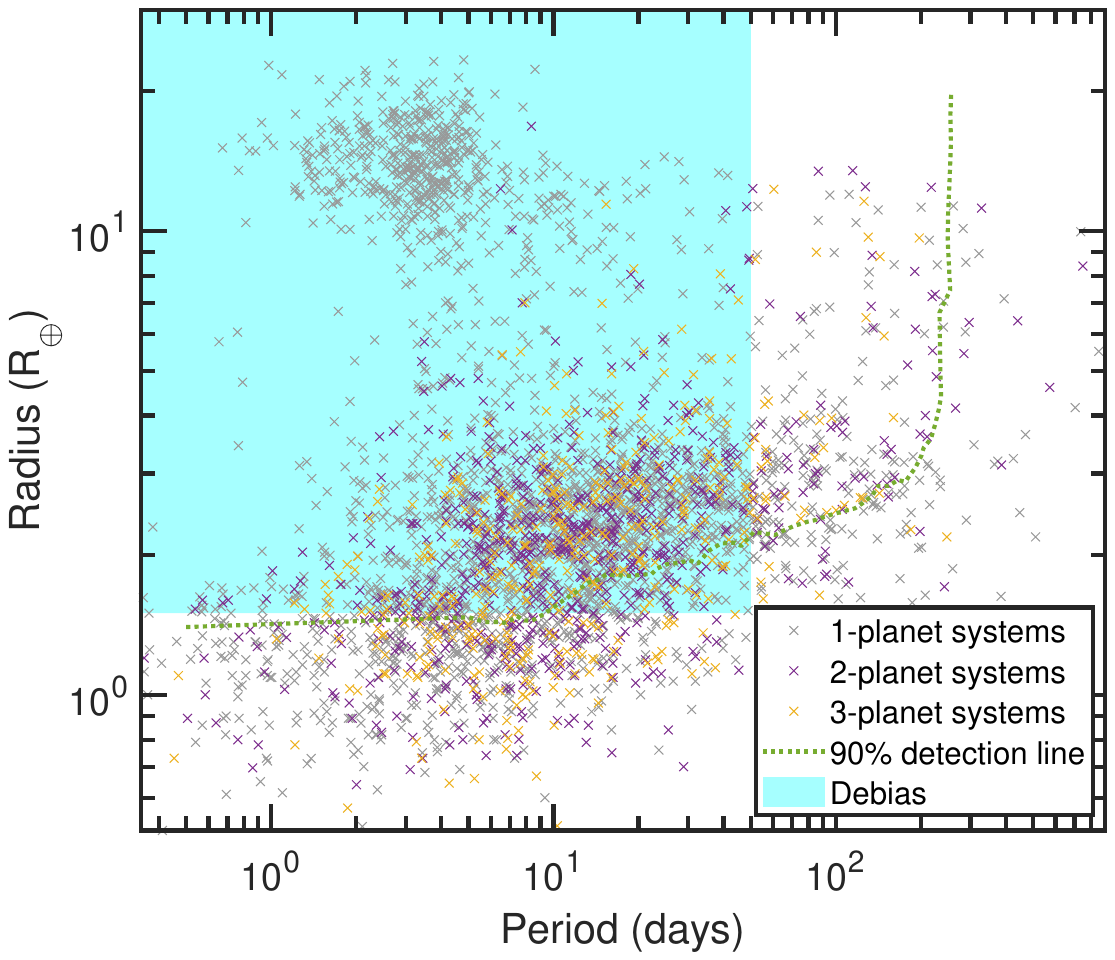}
	\caption{Sample of two- and three-planet systems used in this study. The dashed green line is a 90\% detection threshold of \textit{Kepler}, based on \citet{Petigura2013},{{the shaded area is area that is considered to suffer less from Kepler selection effect, and related as "debiased" later.}} See sections \ref{subsec:Sample} and \ref{subsec:Class}.}
	\label{fig:sampl}
\end{figure}

The planetary sample we use in this study is based on the online publicly available \textit{NASA} Exoplanet Archive \citep{NASA_Exoplanet}. We select planets with measured radius and orbital period. This wide and heterogeneous sample includes planets detected by different missions and suffers from different selection effects.% Later we refer to a sample of planets where the data incompleteness was not taken into account as the "full" sample. 

\citet{Petigura2013} investigated the prevalence of Earth-sized planets orbiting Sun-like stars, considering \textit{Kepler}’s detection limitations and survey incompleteness. They identified a region in Radius-Period space where planets have more than a 90\% chance of being detected by \textit{Kepler}'s CCD. In this study, we use \citet{Petigura2013}'s findings to create a new sub-sample by selecting planets detected by \textit{Kepler} that exceed the 90\% detectability threshold. This approach eliminates planets within the Period-Radius region that may be affected by data incompleteness.{{ However, taking planets above 90\% detectability line introduces a correlation between period and radius, and therefore cannot be used as de-baised sample. Instead of selecting planets above the curve, we select planes in a region of R $>$ 2$R_\oplus$ and P $<$ 50 days (shaded area in figure 1). The planets in the shaded area are refereed later as "de-biased" sample, though it may still experience some minor selection effects, while the sample that does not take any selection effect into account is referred to as "full" }}
%Hoverer, taking planets above 90\% lines We refer to this sample of planets above the 90\% detection line as the "de-biased" sample, though it may still experience some minor selection effects, while the sample that does not take any selection effect into account is referred to as "full".

The "full" (biased) sample of planets is presented in Figure \ref{fig:sampl}, and the dashed green line represents 90\% detection line from \citet{Petigura2013}. The planet above the dashed green line is part of the "de-biased" sample. The classification of two- and three-planet samples is explained in section \ref{subsec:Class}.

%One may notice few planets below 90\% \textit{Kepler}'s detection line: those planets were detected by instruments other than \textit{Kepler}'s CCD. The division to sub-samples is described in section \ref{subsec:Class}. 

\subsection{System classification} 
\label{subsec:Class}

The sample is initially categorized by the host star types of the planetary systems (see later in section \ref{sec:stellar}). Within each stellar type category, the sample is further divided based on the number of planets in each system. Additionally, we analyze an entire planetary sample where systems are only subdivided by the number of planets, without considering the host star types.

We classify each system by the number of observed planets surrounding a given star. The counting of planets per system is done after filtering out planets without measured orbital periods \textit{P} and planetary radii \textit{R}. In the de-biased sample, the classification of systems to two- and three-planet systems is done after filtering out planets with {{R $>$ 2$R_\oplus$ and P $<$ 50 days, and therefore some planetary systems have different classifications in full and de-biased samples, (See Figure \ref{fig:systems}), as some systems change their classification after debiasing. Therefore, the "debiased" 2-planet sample is not exactly a subset of the "full" 2-planet sample. This is due to the fact that a system originally classified as a 3-planet system may be reclassified as a 2-planet system after applying the threshold filter. De-biased sample might include some systems that were not counted in corresponding "full" sample.}}

\begin{table}[h]
\centering
\begin{tabular}{|c|c|c|c|c|c|}
\hline
 & 12 & 21 & Total \\
\hline
Full & 273 (74.59\%) & 93 (25.41\%) & 366 \\
De-biased & 153 (69.23\%) & 68 (30.77\%) & 221 \\
\hline
\end{tabular}
\caption{Contingency table for "full" and "debiased" 2-planet samples' configurations.}\label{tbl:2pln}
\end{table}

\begin{table}[h]
\centering
\begin{tabular}{|c|c|c|c|c|c|c|c|}
\hline
 & 123 & 132 & 213 & 231 & 312 & 321 & Total \\
\hline
Full & 57 (41.61\%) & 38 (27.74\%) & 16 (11.68\%) & 13 (9.49\%) & 9 (6.57\%) & 4 (2.92\%) & 137 \\
De-biased & 21 (31.82\%) & 20 (30.30\%) & 6 (9.09\%) & 5 (7.58\%) & 10 (15.15\%) & 4 (6.06\%) & 66 \\
\hline
\end{tabular}
\caption{Contingency table for "full" and "debiased" 3-planet samples' configurations. For pairs out of the three, see table \ref{tbl:pairs_out_three}.}\label{tbl:3pln}
\end{table}

\begin{table}[h]
\centering
\begin{tabular}{|c|c|c|c|}
\hline
 & 12 & 21 & Total \\
\hline
ab/3 Full & 109 (79.56\%) & 28 (20.44\%) & 137 \\
ab/3 De-biased & 46 (69.70\%) & 20 (30.30\%) & 66 \\
ac/3 Full & 104 (75.91\%) & 33 (24.09\%) & 137 \\
ac/3 De-biased & 42 (63.64\%) & 24 (36.36\%) & 66 \\
bc/3 Full & 88 (64.23\%) & 49 (35.77\%) & 137 \\
bc/3 De-biased & 42 (63.64\%) & 24 (36.36\%) & 66 \\
\hline
\end{tabular}
\caption{Contingency table for "full" and "debiased" configurations for pairs of 3-planet systems. {{"a", "b" and "c" are innermost, middle and outermost planets in each system. "ab"/3 stands for "the innermost two planets out of three". See section \ref{subsec:Class}. } }}\label{tbl:pairs_out_three}
\end{table}

\begin{figure}[h]
	\centering

	\includegraphics[width=1\columnwidth,clip]{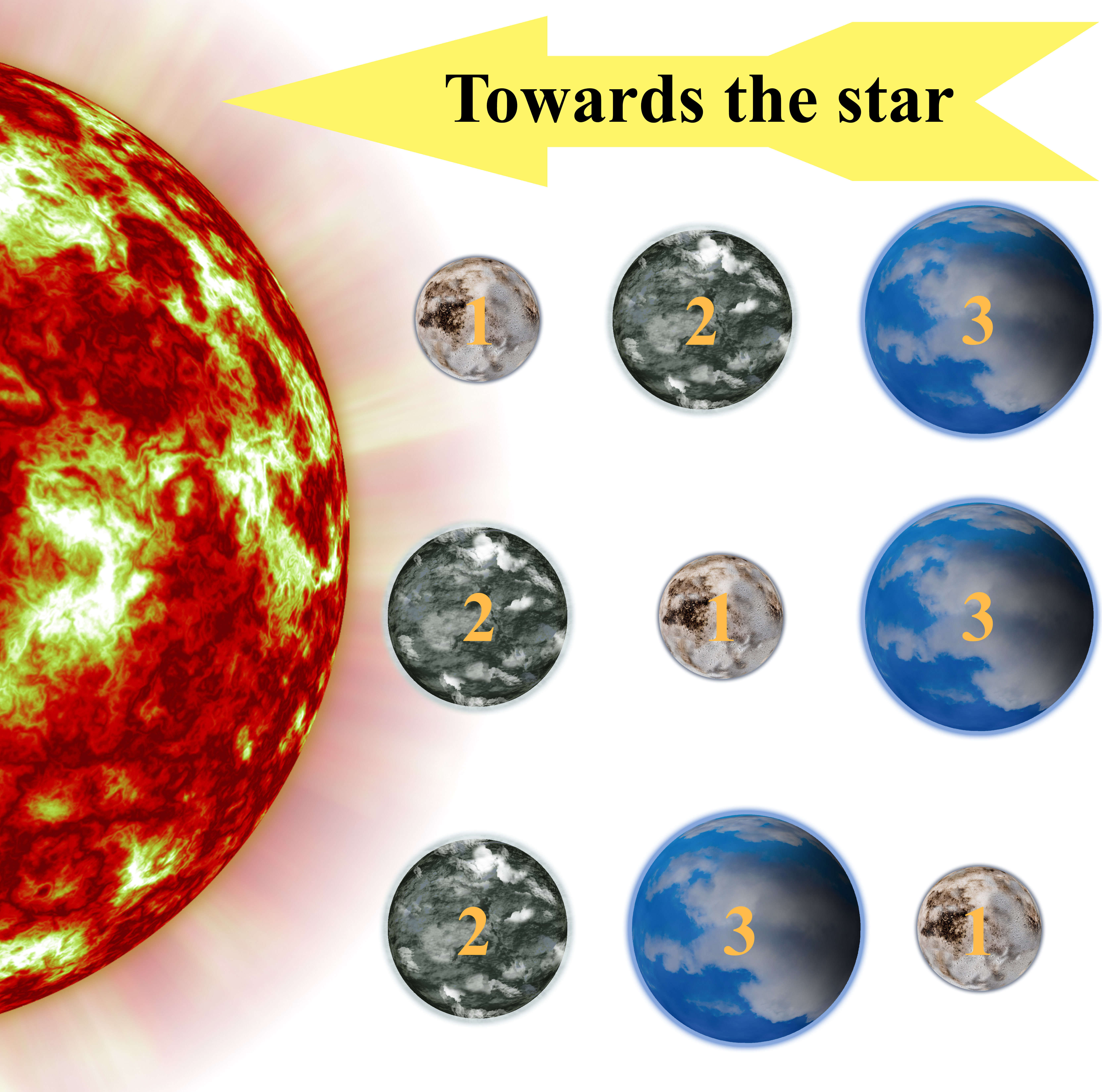}
	\caption{A planetary system is classified by the order of its planets' relative sizes, counting from the innermost one outwards. Here we schematically present three types of systems (out of the possible six configurations): 123; 213 and 231.  }
	\label{fig:order}
\end{figure}

\begin{figure}[h]
	\centering

	\includegraphics[width=1\columnwidth, trim={3.0cm 6.0cm 4.0cm 6.5cm},clip]{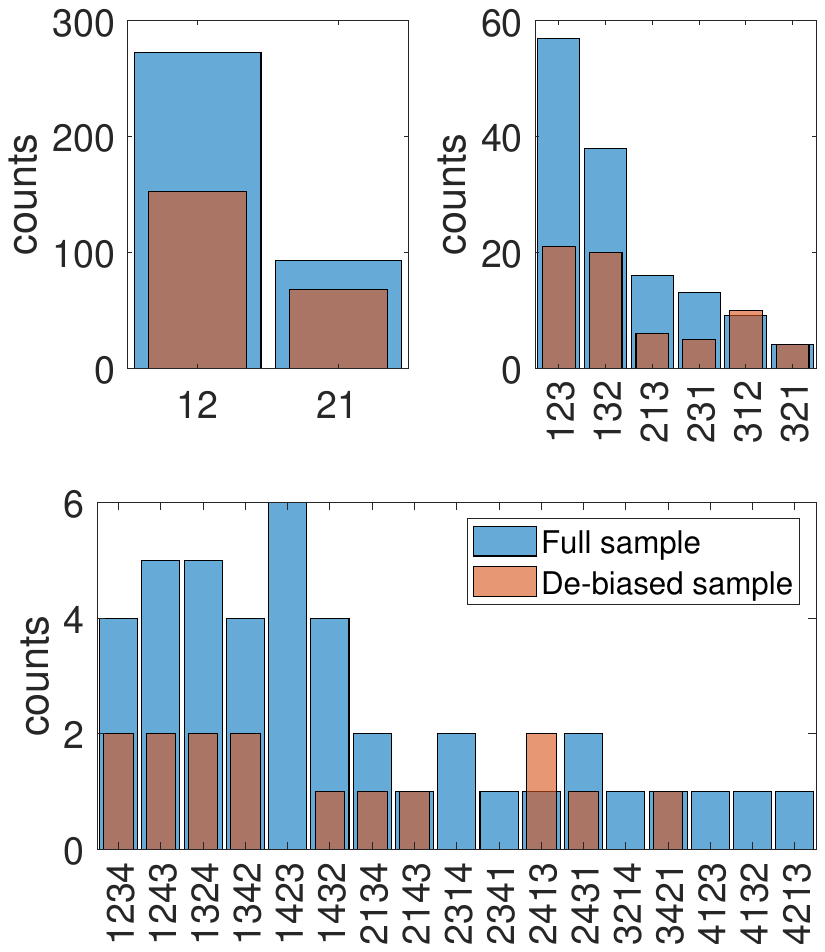}
	\caption{The distribution of planetary ordering for two- (upper left), three- (upper-right) and four-planet(lower) systems. The histogram shows the full sample versus de-biased samples.{{The corresponding contingency tables for two-planet and three-planet systems' sample are listed in Tables \ref{tbl:2pln} and \ref{tbl:3pln}. Four-planet systems were not studied, due to low numbers of planets in each configuration.}} }
	\label{fig:systems}
\end{figure}

After categorizing systems by the number of planets they contain, each two-, three-, and four-planet system is then sub-classified based on the sequence of its relative planetary sizes, starting from the star outward. For example: a system with two planets where the inner one is smaller than the outer one, will be called "12"; a system with two planets where the inner one is larger than the outer one, will be called "21"; In a three-planet system if the largest planet is the innermost, and the smallest one is next, then the system will be called "312". See Figure \ref{fig:order} for three examples of planetary system structures{{and Tables \ref{tbl:2pln}-\ref{tbl:3pln} for quantities of two- and three-planet systems }}.

%\begin{table}[h]\caption {The samples divided to several types of systems. "No." stands for the number of planets in the system, while the numbers show the total number of systems in the full and the de-biases samples. }
%\begin{tabular}{|l|l|l|}
%\hline
%No. & Full & De-biased \\ \hline
%2   & 379  & 246       \\ \hline
%3   & 143  & 70        \\ \hline
%4   & 46   & 26        \\ \hline
%5   & 19   & 2         \\ \hline
%\end{tabular}\label{tbl:sample}
%\end{table}

%Two-, three- and four-planet systems divided into their sub-classes are presented in Figure \ref{fig:systems}. We present the counts of each sub-class. In this stage it seems like the inner planets tend to be smaller than the outer ones. This hypothesis is tested later in section \ref{subsc:KS}. Due to relative low number of four-, and five-planet  systems (not presented) we focus on two- and-three planet systems.

%As a next step, we focus on three-planet systems. We select planetary pairs out of three and examine their relative sizes. We denote pairs out of three-planet systems by their relative orbital periods, meaning that the inner two planets in the three-planet system will be called "ab/3". The outer two planets will be called "bc/3" and the pair of the innermost one and the outermost one out of three will be called "ac/3". Note that this notation relates to orbital location only, without relations to the relative sizes of planets. In this manner, the last two planets out of three (bc/3) might be in a configuration of "12" or "21", meaning that the first one is smaller or the first one is larger correspondingly. The counts of pairs out of three are shown in Figure \ref{fig:Subsystems}.  In few cases one may notice that there are more three-planet cases of some configuration in the de-biased sample compared to the full one. This results from reducing four-planet systems to three-planet systems by excluding planets which reside below the 90\% confidence limit.

Two-, three-, and four-planet systems subdivided into their relative sizes configurations are presented in Figure \ref{fig:systems}. We present the counts of each configuration. It seems that the inner planets tend to be smaller than the outer ones. This hypothesis is tested later in section \ref{subsc:KS}. Due to the relatively low number of four- and five-planet systems (not shown), we focus on two- and three-planet systems only.

\begin{figure}[h]
	\centering

	\includegraphics[width=1\columnwidth, trim={2.0cm 8.5cm 3.0cm 7.0cm},clip]{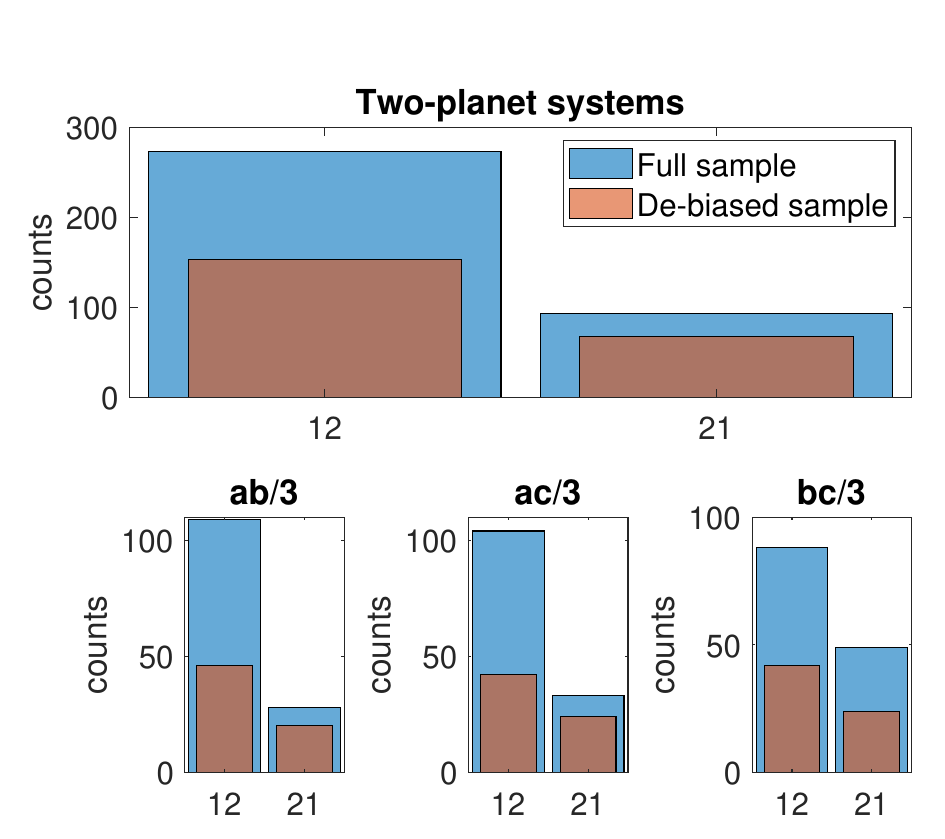}\caption{Two planet systems compared to pair out of three planets systems.{{Corresponded contingency table is Table \ref{tbl:pairs_out_three}.}}}
	\label{fig:Subsystems}
\end{figure}

As the next step, we focus on three-planet systems. We select planetary pairs out of these systems and examine their relative sizes. We denote pairs out of three-planet systems by their relative orbital periods: the inner two planets are called "ab/3", the outer two planets "bc/3", and the pair of the innermost and outermost planets "ac/3". This letter notation relates to orbital location only, without relation to the relative sizes of the planets: Letters refers to the relative orbital location in a system, and numbers to relative sizes of planets.  Thus, the last two planets out of three "bc/3" (for example) could be in a configuration of "12" if the first is smaller, or "21" if the first is larger. In this way, each three-planet system is studied as three two-planet subsystem configurations{{( ab/3, ac/3, bc/3)}}. The counts of pairs out of three-planet systems are shown in Figure \ref{fig:Subsystems}{{and Table \ref{tbl:pairs_out_three}}}. One may note that in some cases, there are more three-planet configurations in the "de-biased" sample compared to the "full" one. This results from reducing four-planet systems to three-planet systems by excluding planets below our confidence limit.

%We divide the three-planet systems to sub-systems, by removing innermost, middle or outermost planet in each system. Those systems are referred as "1,2","1,3" and "2,3" and presented in Figure \ref{fig:Subsystems}.

\subsection{Stellar type}
\label{sec:stellar}

\begin{figure}[h]
	\centering

	\includegraphics[width=1\columnwidth, trim={0.7cm 6.4cm 1.7cm 6.0cm},clip]{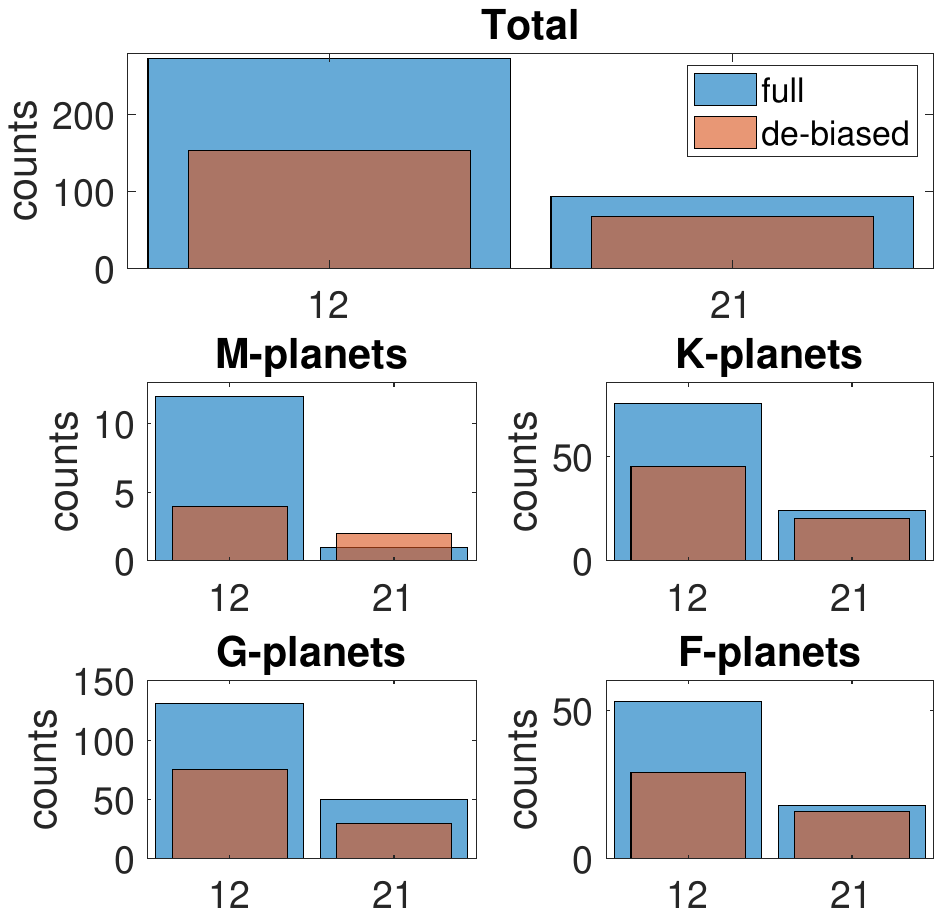}
	\caption{The ordering distribution of two-planet systems. The full sample and the de-biased ones are divided into stellar types. {{Corresponding contingency table is Table \ref{tbl:2plnStars}}}}
	\label{fig:stars2}
\end{figure}

\begin{figure}[h]
	\centering

	\includegraphics[width=1\columnwidth, trim={1.7cm 6.5cm 2.5cm 6.0cm},clip]{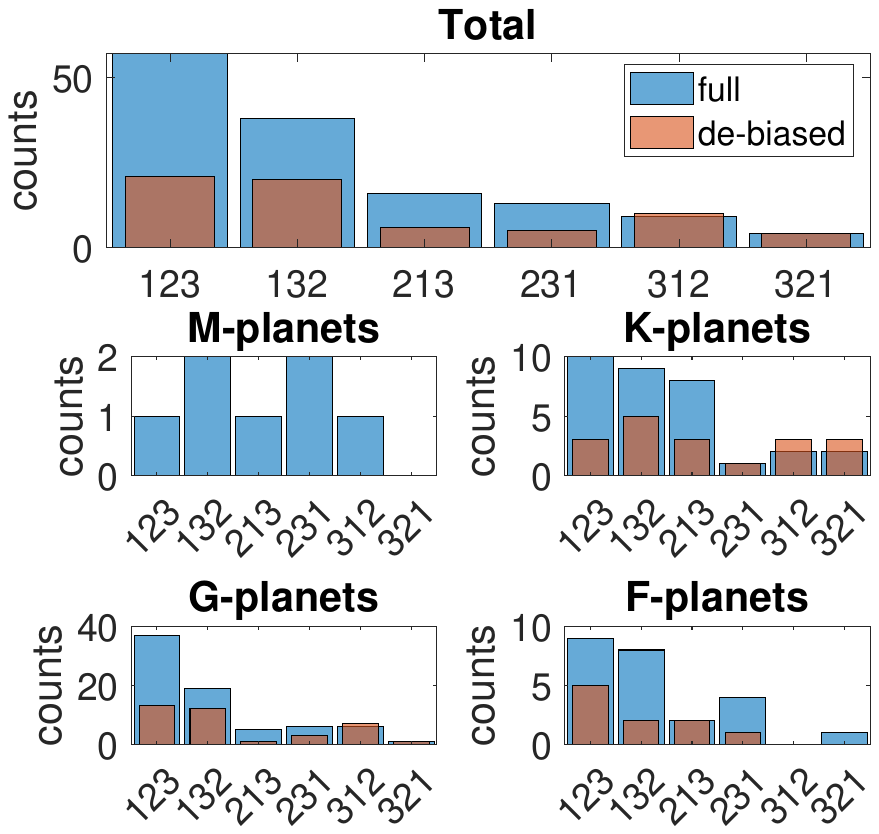}
	\caption{The ordering distribution of three-planet systems. The full sample and the de-biased ones are divided into stellar types. {{Corresponding contingency table is Table \ref{tbl:3plnStars}}} }
	\label{fig:stars3}
\end{figure}

It was found that planetary sizes scale with the stellar type, a trend that cannot be explained by data incompleteness alone \citep{Lozovsky2021}. We can divide our sample by the stellar type of the systems'{{hosts}}, where the stellar type itself was calculated from the stellar temperature. The two-planet systems' histograms with the stellar classifications are presented in Figure \ref{fig:stars2}, and the three-planet systems are presented in Figure \ref{fig:stars3}.

\begin{table}[h]
\centering
\begin{tabular}{|c|c|c|c|}
\hline
 & 12 & 21 & Total \\
\hline
M-planets Full & 12 (92.31\%) & 1 (7.69\%) & 13 \\
M-planets De-biased & 4 (66.67\%) & 2 (33.33\%) & 6 \\
K-planets Full & 75 (75.76\%) & 24 (24.24\%) & 99 \\
K-planets De-biased & 45 (69.23\%) & 20 (30.77\%) & 65 \\
G-planets Full & 131 (72.38\%) & 50 (27.62\%) & 181 \\
G-planets De-biased & 75 (71.43\%) & 30 (28.57\%) & 105 \\
F-planets Full & 53 (74.65\%) & 18 (25.35\%) & 71 \\
F-planets De-biased & 29 (64.44\%) & 16 (35.56\%) & 45 \\
\hline
\end{tabular}
\caption{Contingency table for two-planet systems, divided to different stellar types. }\label{tbl:2plnStars}

\end{table}

\begin{table}[h]
\centering
\begin{tabular}{|c|c|c|c|c|c|c|c|}
\hline
 & 123 & 132 & 213 & 231 & 312 & 321 & Total \\
\hline
M-planets Full & 1 (14.29\%) & 2 (28.57\%) & 1 (14.29\%) & 2 (28.57\%) & 1 (14.29\%) & 0 (0.00\%) & 7 \\
M-planets De-biased & 0 (-) & 0 (-) & 0 (-) & 0 (-) & 0 (-) & 0 (-) & 0 \\
K-planets Full & 10 (31.25\%) & 9 (28.12\%) & 8 (25.00\%) & 1 (3.12\%) & 2 (6.25\%) & 2 (6.25\%) & 32 \\
K-planets De-biased & 3 (16.67\%) & 5 (27.78\%) & 3 (16.67\%) & 1 (5.56\%) & 3 (16.67\%) & 3 (16.67\%) & 18 \\
G-planets Full & 37 (50.00\%) & 19 (25.68\%) & 5 (6.76\%) & 6 (8.11\%) & 6 (8.11\%) & 1 (1.35\%) & 74 \\
G-planets De-biased & 13 (35.14\%) & 12 (32.43\%) & 1 (2.70\%) & 3 (8.11\%) & 7 (18.92\%) & 1 (2.70\%) & 37 \\
F-planets Full & 9 (37.50\%) & 8 (33.33\%) & 2 (8.33\%) & 4 (16.67\%) & 0 (0.00\%) & 1 (4.17\%) & 24 \\
F-planets De-biased & 5 (50.00\%) & 2 (20.00\%) & 2 (20.00\%) & 1 (10.00\%) & 0 (0.00\%) & 0 (0.00\%) & 10 \\
\hline
\end{tabular}
\caption{Same as \ref{tbl:2plnStars}, for three-planet systems.}\label{tbl:3plnStars}
\end{table}

We should note that the vast majority of the planets surrounding M stars are below \citet{Petigura2013} 90\% detection limit and are therefore absent from the de-biased sample.

\subsection{Planetary Type}

\begin{figure}[h]
	\centering

	\includegraphics[width=1\columnwidth, trim={2.0cm 8.0cm 2.8cm 7.4cm},clip]{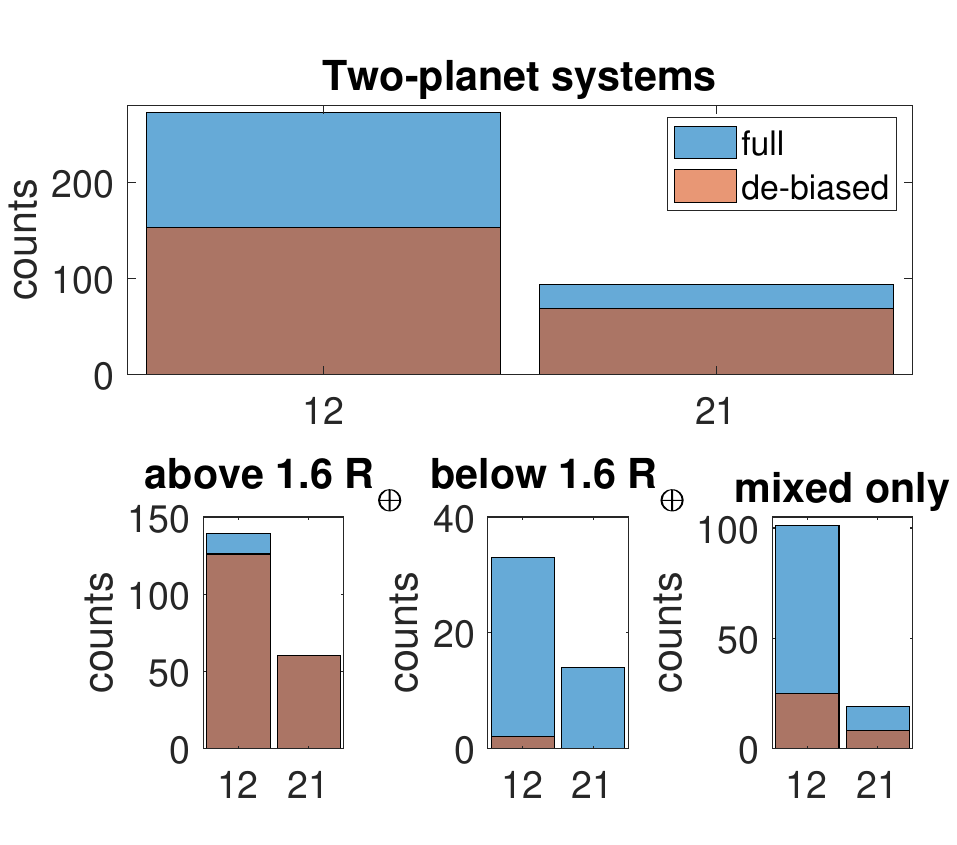}
	\caption{Two-planet systems divided into sub-groups by the planetary sizes: all the planets are above 1.6 R$_\oplus$, all the planets are below R$_\oplus$, or mixed.{{The corresponding contingency table is Table \ref{tbl:2pln1.6}}} }
	\label{fig:planet2}
\end{figure}

\begin{figure}[h]
	\centering

	\includegraphics[width=1\columnwidth, trim={2.0cm 7.9cm 2.8cm 7.4cm},clip]{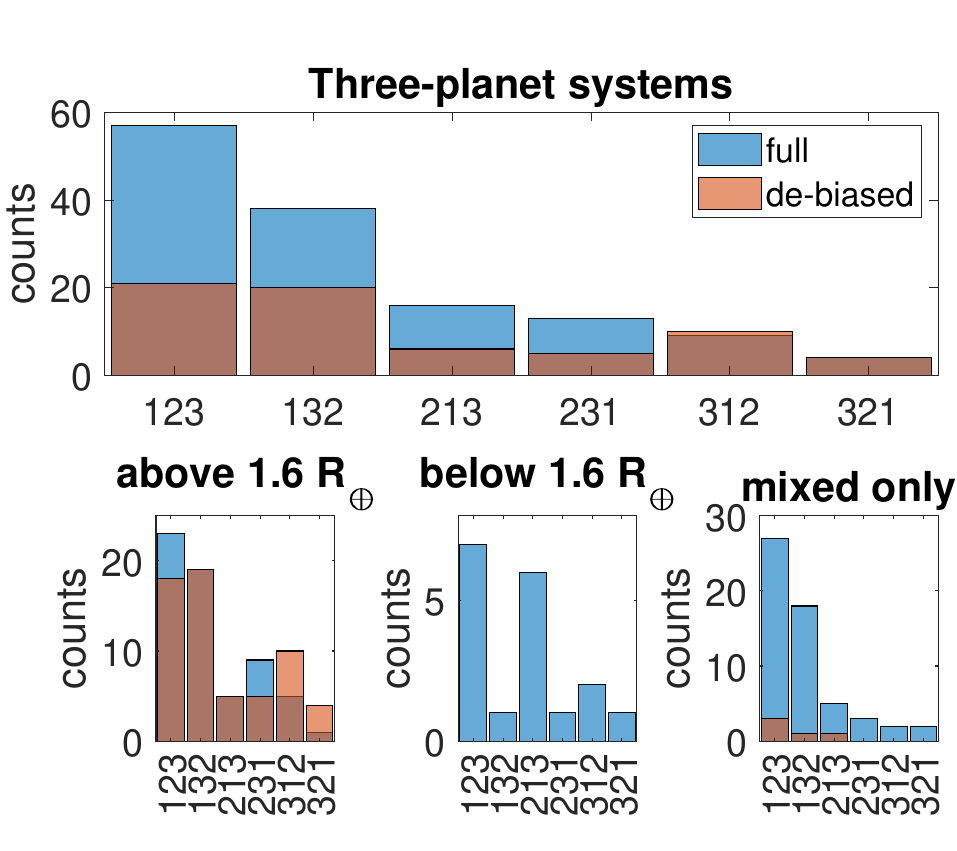}
	\caption{Same as Figure \ref{fig:planet2}, for three-planet systems.{{The corresponding contingency table is Table \ref{tab:3pln1.6}}} }
	\label{fig:planet3}
\end{figure}

It was found that planets with radii above 1.6 R$_\oplus$ are most probably not rocky \citep{Rogers2015,Lozovsky2018,Mousis2020}. Therefore we use this radius threshold to distinguish between potentially rocky and potentially non-rocky populations. Two- and three-planet systems are divided into three categories: systems composed only of planets that are below the 1.6 R$_\oplus$ threshold; systems composed only of planets above the threshold; and systems that have a least one planet of each type ("mixed"). The systems are presented in figures \ref{fig:planet2} and \ref{fig:planet3}.

\begin{table}[h]
\centering
\begin{tabular}{|c|c|c|c|}
\hline
 & 12 & 21 & Total \\
\hline
Above 1.6 R$_\oplus$ Full & 139 (69.85\%) & 60 (30.15\%) & 199 \\
Above 1.6 R$_\oplus$ De-biased & 126 (67.74\%) & 60 (32.26\%) & 186 \\
Below 1.6 R$_\oplus$ Full & 33 (70.21\%) & 14 (29.79\%) & 47 \\
Below 1.6 R$_\oplus$ D-ebiased & 2 (100.00\%) & 0 (0.00\%) & 2 \\
Mixed Full & 101 (84.17\%) & 19 (15.83\%) & 120 \\
Mixed De-biased & 25 (75.76\%) & 8 (24.24\%) & 33 \\
\hline
\end{tabular}
\caption{Contingency table for two-planet systems, divided to planetary classes by planetary sizes: both of the planets are above 1.6 R$_\oplus$, both are below 1.6 R$_\oplus$, or one of each type.}\label{tbl:2pln1.6}
\end{table}

\begin{table}[h]
\centering
\begin{tabular}{|c|c|c|c|c|c|c|c|}
\hline
 & 123 & 132 & 213 & 231 & 312 & 321 & Total \\
\hline
Above 1.6 R$_\oplus$ Full & 23 (37.10\%) & 19 (30.65\%) & 5 (8.06\%) & 9 (14.52\%) & 5 (8.06\%) & 1 (1.61\%) & 62 \\
Above 1.6 R$_\oplus$ De-biased & 18 (29.51\%) & 19 (31.15\%) & 5 (8.20\%) & 5 (8.20\%) & 10 (16.39\%) & 4 (6.56\%) & 61 \\
Below 1.6 R$_\oplus$ Full & 7 (38.89\%) & 1 (5.56\%) & 6 (33.33\%) & 1 (5.56\%) & 2 (11.11\%) & 1 (5.56\%) & 18 \\
Below 1.6 R$_\oplus$ De-biased & 0 (-) & 0 (-) & 0 (-) & 0 (-) & 0 (-) & 0 (-) & 0 \\
Mixed Full & 27 (47.37\%) & 18 (31.58\%) & 5 (8.77\%) & 3 (5.26\%) & 2 (3.51\%) & 2 (3.51\%) & 57 \\
Mixed De-biased & 3 (60.00\%) & 1 (20.00\%) & 1 (20.00\%) & 0 (0.00\%) & 0 (0.00\%) & 0 (0.00\%) & 5 \\
\hline
\end{tabular}
\caption{same as table \ref{tbl:2pln1.6}, for three-planet systems}
\label{tab:3pln1.6}
\end{table}

Figure \ref{fig:planet2} indicates that most of the major contributions to the trend where the inner planet is smaller than the outer one come from planets above 1.6 R$_\oplus$, which are not rocky. For systems where both of the planets are rocky, the opposite trend can be seen: "21" configuration is much more common than "12". As the majority of the planets in the de-biased sample are not rocky, the result might be due to selection bias.

\section{Results}
\label{sec:results}

\begin{figure}[h]
	\centering

	% Top panel: CDF
	\includegraphics[width=0.85\columnwidth, trim={0.0cm 4.5cm 0.0cm 4.6cm}, clip]{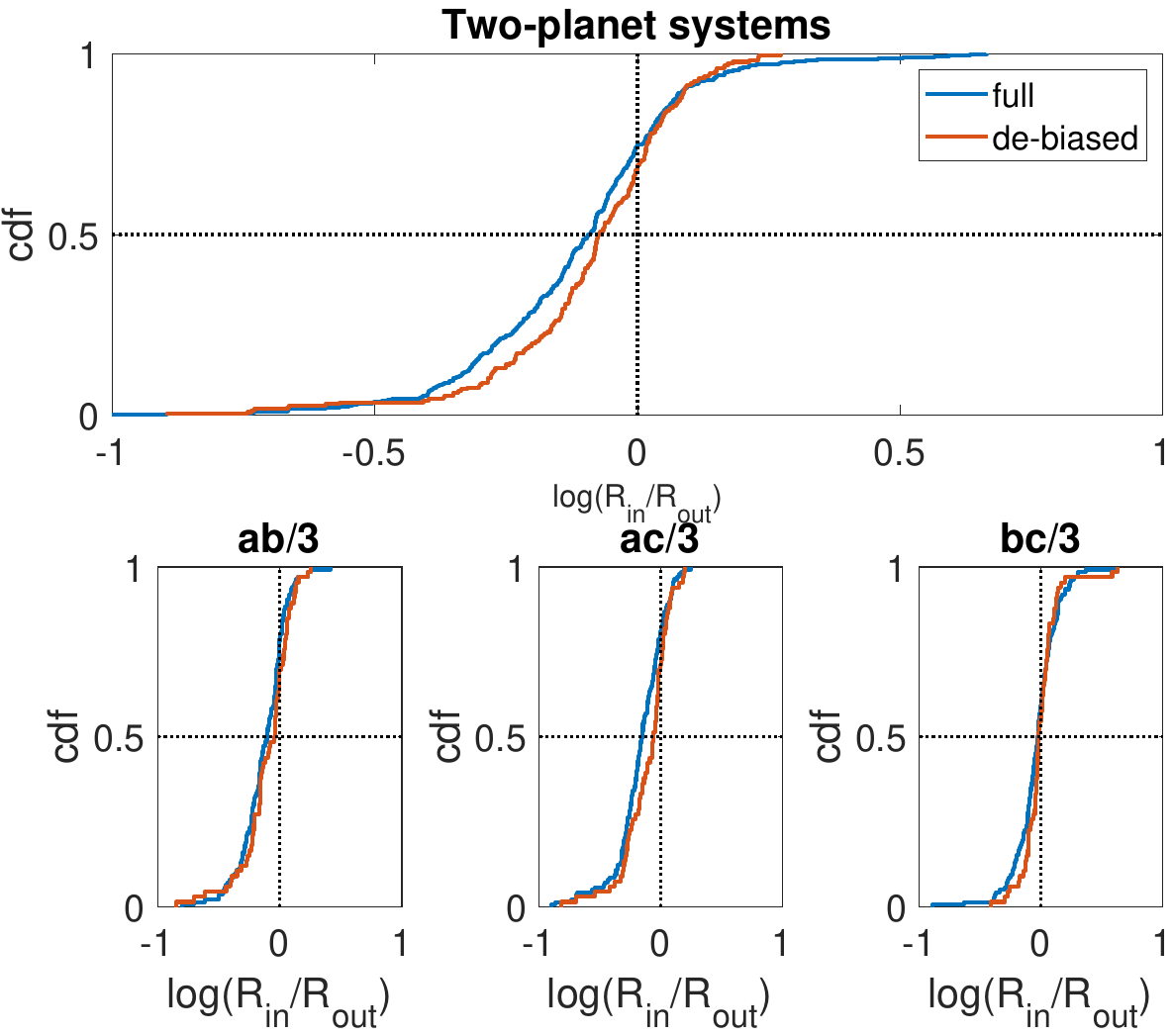}
	
	\vspace{0.005cm} % small space between panels

	% Bottom panel: PDF
	\includegraphics[width=0.85\columnwidth, trim={0.0cm 9.5cm 0.5cm 10.0cm}, clip]{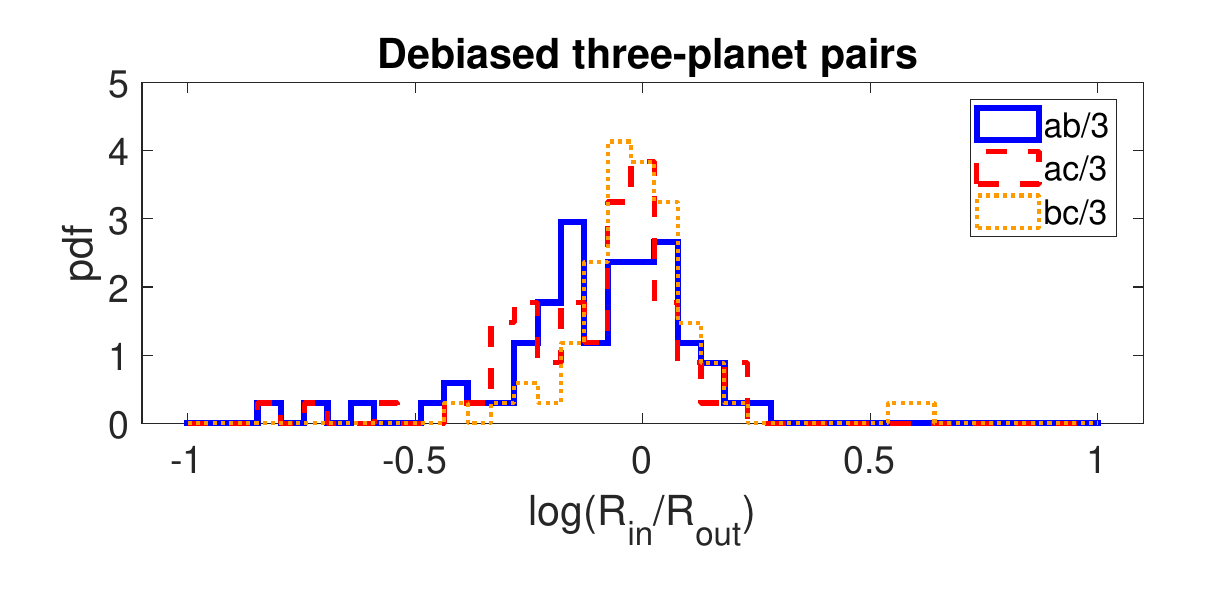}
	
	\caption{Top: Cumulative distribution function (CDF) of the ratio of inner to outer planets' radii for two-planet systems. Middle: same for two-planet pairs in three-planet systems. Bottom: Probability density function (PDF) of the "debiased" samples as in the middle. System classifications are described in section~\ref{subsec:Class}.}
	\label{fig:CDF_PDF}
\end{figure}

The tests indicate that the inner planets in multi-planet systems tend to have smaller radii compared with outer planets in the same systems. The Cumulative Distribution Function (CDF) of the ratio $R_{in}/R_{out}$ is presented in Figure \ref{fig:CDF_PDF}. {{This Figure indicates an interesting trend where the outer pair out of three (bc/3) tends to have similar radii, following `peas in a pod' hypothesis, whereas the other configurations do not.}}

%In can be seen that the last two planets in three-planet systems seem to have very different distribution compared to two-planet systems (see also Figure \ref{fig:Subsystems}). It is interesting to note that the behavior of the three-planet systems shows differences between the full-sample and the de-biased one. This can be seen in Figures \ref{fig:systems} and \ref{fig:stars3}. In contrary to the de-biased sample the combination "312" is more abundant than "321" and "231", which may indicate that the main trend might arise from some selection effect. 

%We find that a noticeable divergence arises in the distributions of the last two planets within three-planet systems compared to those within two-planet systems (see also Figure \ref{fig:Subsystems}). Of particular interest is the contrast observed between the behaviors of three-planet systems within the full dataset and those within the de-biased subset. This contrast is also illustrated in Figures \ref{fig:systems} and \ref{fig:stars3}. Notably, contrary to the de-biased subset, the "312" configuration appears more prevalent than "321" and "231", indicating a potential influence of selection biases on the observed trend. {{However, given low number of planets in each bin (5-6), this conclusion must be taken with caution.}}

%In the three-planet systems we can note that the behavior of "12" and "13" and this of "23". 
Figures \ref{tbl:pairs_out_three}-\ref{fig:planet3} and Tables \ref{tbl:2pln}-\ref{tab:3pln1.6} indicate that configurations where the inner planets are smaller are more common. A quantitative comparison of systems is done in Section \ref{stats}.
Figure \ref{fig:Subsystems} suggests that in both the de-biased and full samples, the probability for the outermost planet to be the largest is significantly lower compared to other configurations, i.e. the ordering of the inner planet being the smallest is mostly true for first two planets and less so for outer pairs (see also four-planet systems in Figure \ref{fig:systems}). This can be seen also in Figure \ref{fig:CDF_PDF}, where the CDFs of the two-planet samples, ab/3 and ac/3 look similar, but bc/3 seems to be different (tested later in Section \ref{stats} and Table \ref{tab:KS}).

%The two-sample K-S test (see section \ref{subsc:KS}) could not reject a hypothesis that the innermost two planets in three-planet systems (ab/3) come from the same distribution as two-planet systems (Table \ref{tab:KS}). The same could be said about the first and last out of three (ac/3) in the de-biased sample. But the bc/3 distribution seems to be significantly different. 
%For every pair, the test did not reject that the de-biased sample comes from the same distribution as the full one, as the de-biased is mostly a subset of the full one. 
%, suggesting that possibly the "two-planet systems" are expected to be extended to three-planet systems, with the new discoveries of outer planets. For every pair the test did not reject that the de-biased sample comes from the same distribution as the full one, as the de-biased is mostly a sub-set of the full one. 

\subsection{Orbital resonances}

Several groups \citep[e.g.][]{Dawson2019,Dawson2021,Charalambous2022}{{have pointed out that the planetary orbits of exoplanets show a lack of planets with orbital period ratios close to integer numbers, which are typically associated with orbital resonances}} \citep[e.g.][]{Armstrong2015,Dawson2019,Leleu2021}.

\begin{figure*}
	\centering

	\includegraphics[width=1\columnwidth, trim={0.0cm 3.0cm 0.0cm 2.0cm},clip]{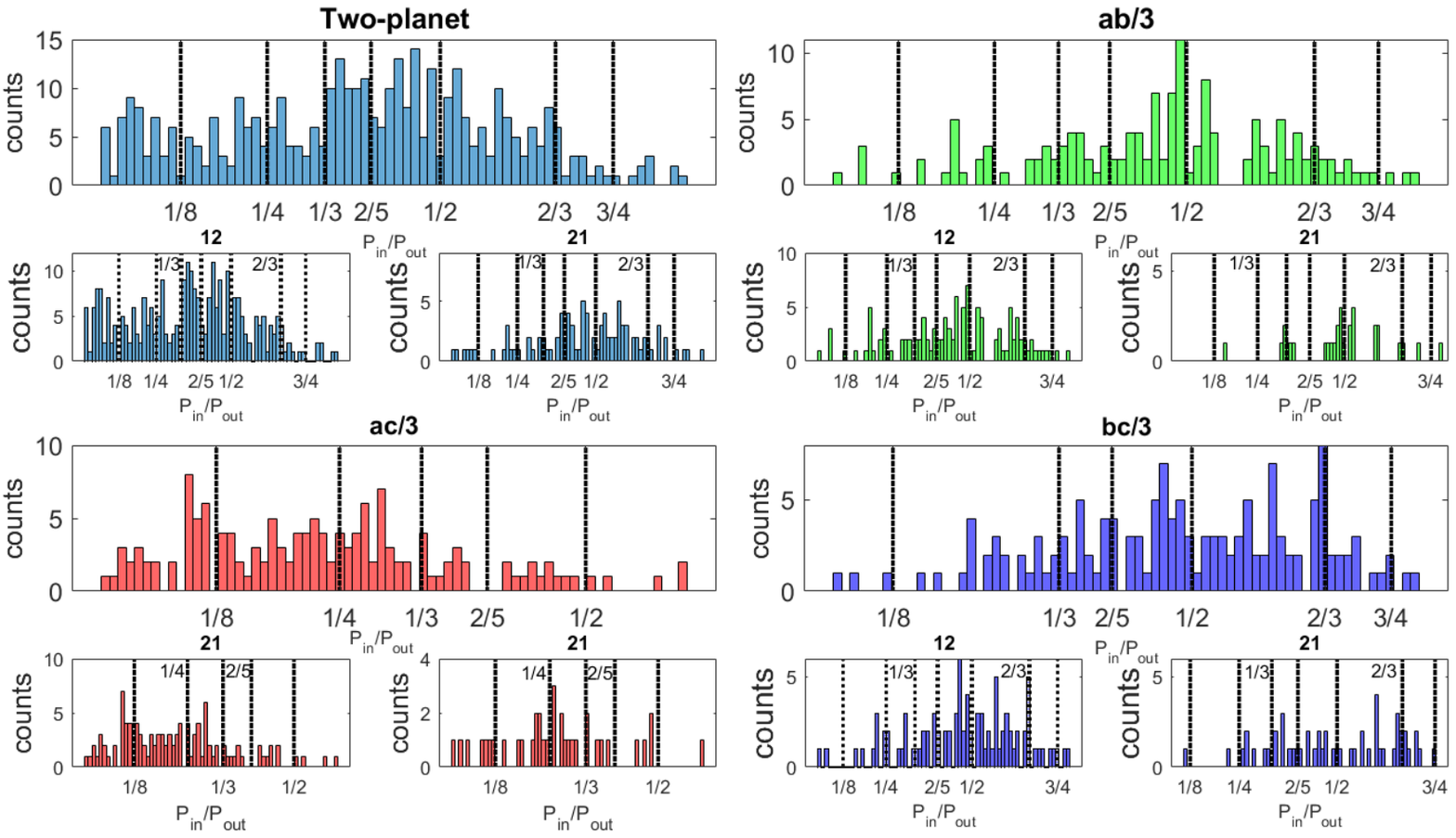}
	\caption{Histograms of inner-to-outer planet ratios $R_{in}/R_{out}$, for two-planet systems, and pairs out of three-planet systems. The dashed lines show integer rations.}
	\label{fig:ResoHist}
\end{figure*}

\begin{figure*}
	\centering

	\includegraphics[width=1\columnwidth, trim={0.0cm 1.0cm 0.0cm 1.0cm},clip]{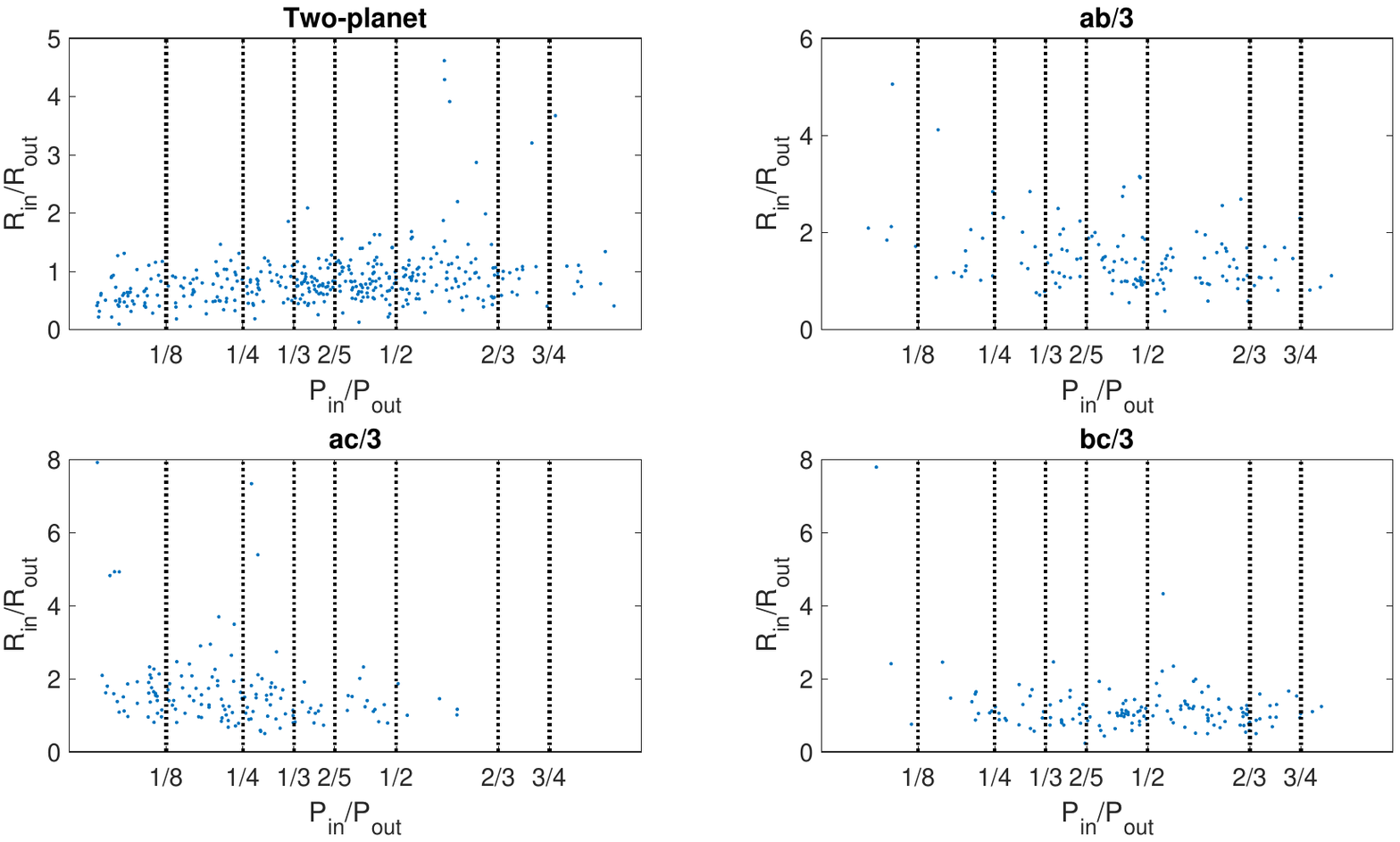}
	\caption{Ratios of planetary radii and periods, for two-planet systems and pairs out of three-planet systems.}
	\label{fig:PinPout}
\end{figure*}

\begin{figure*}
	\centering

	\includegraphics[width=1\columnwidth, trim={0.0cm 2.0cm 0.0cm 2.0cm},clip]{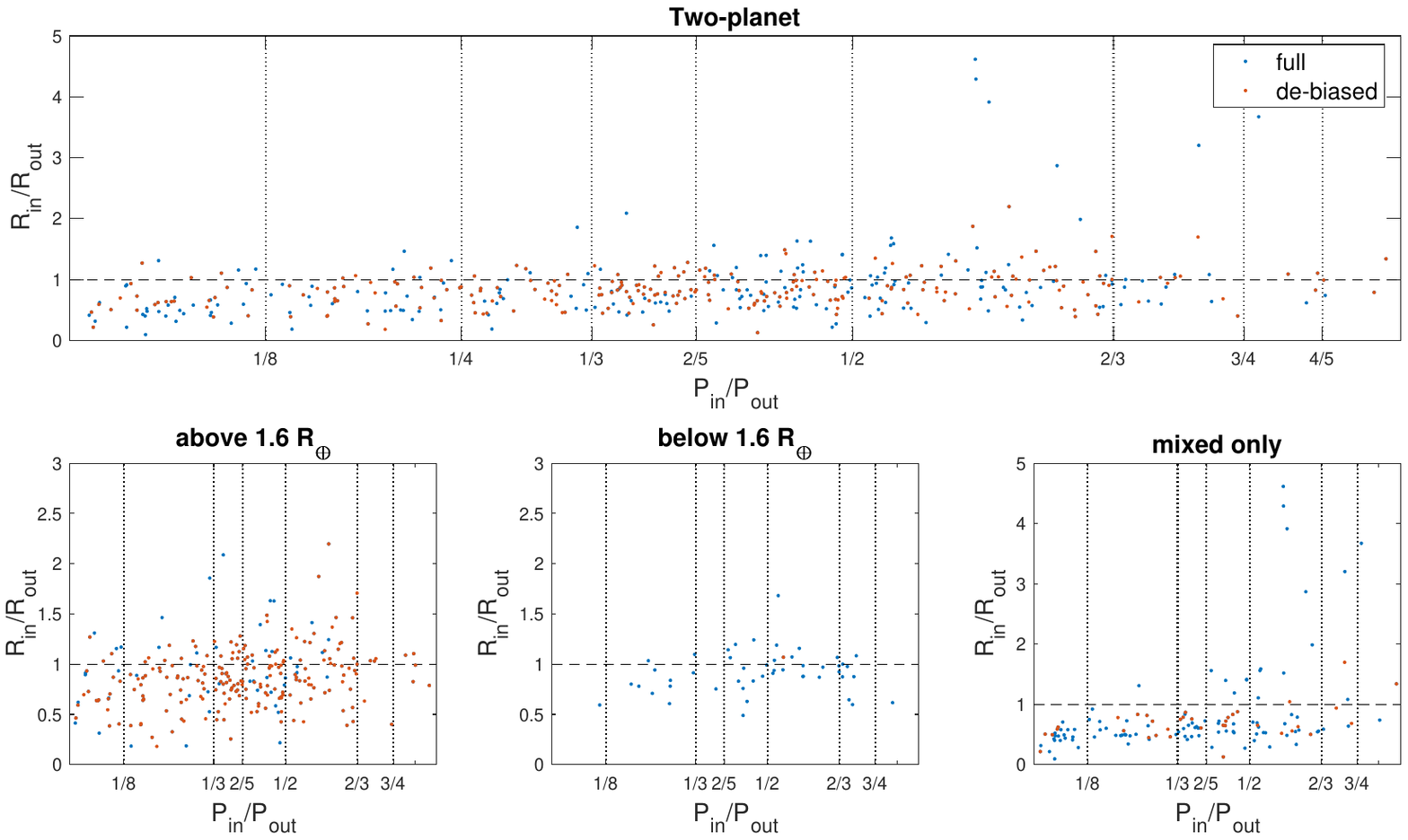}
	\caption{Ratios of planetary radii and periods, for two planet systems and for sub-samples including small planets only, large planets only, or systems with one large planet and one small planet.}
	\label{fig:PinPoutSubSys}
\end{figure*}

 \begin{figure*}
	\centering

	\includegraphics[width=1\columnwidth, trim={3.0cm 0.2cm 3.0cm 0.1cm},clip]{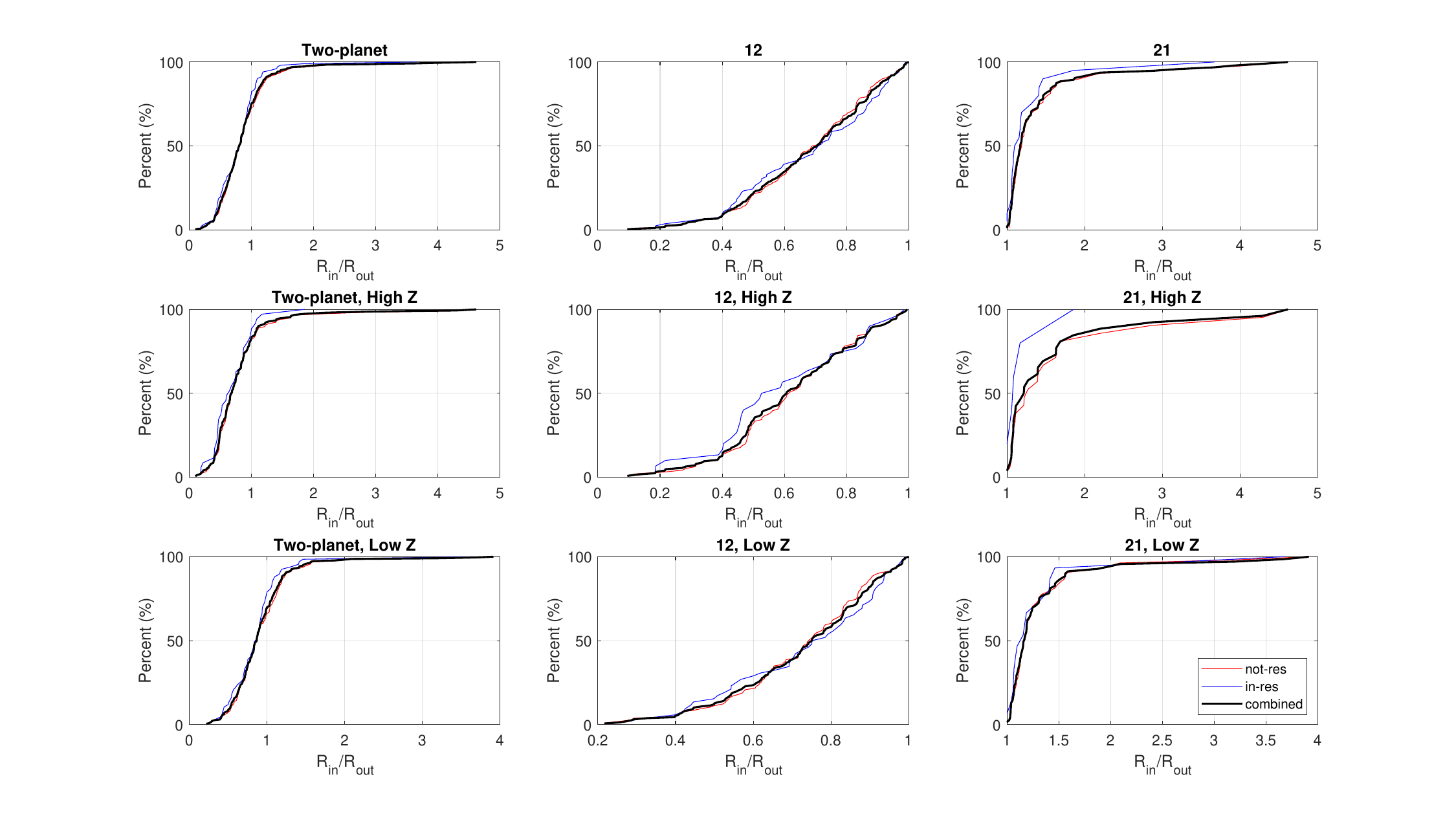}
	\caption{ Normalized cumulative distribution function (CDF) of radius ratios $R_{in}/R_{out}$. Planets with period ratios close to resonances (tolerance of 0.02) are colored blue, those not close to resonances are colored red, and the combined data is colored black. } %Two-sample K-S test did not find statistically significant differences between the sub-samples. }
	\label{fig:CdfRratio}
\end{figure*}

%Figure \ref{fig:ResoHist} shows the ratios of orbital periods of two-planet systems and pairs out of three-planet systems.  In Fig. \ref{fig:CdfRratio} we compare the cumulative distribution function of radius ratios $R_{in}/R_{out}$ of planet pairs close and far from resonances,{{that are indicated by integer rations of periods}}. The distributions in both Fig.   \ref{fig:ResoHist} and \ref{fig:CdfRratio} do not show significant differences between pairs in (or close to) resonance vs. those outside resonances; this is also supported by the KS tests show no significant differences between them.  

Figure \ref{fig:ResoHist} shows the ratios of orbital periods of two-planet systems and pairs within three-planet systems. In Fig. \ref{fig:CdfRratio}, we compare the cumulative distribution function of radius ratios $R_{\mathrm{in}}/R_{\mathrm{out}}$ for planet pairs{{that are close to or far from mean-motion resonances (MMRs), where proximity to resonance is defined as being within 0.02 of a nominal first-order resonance (i.e., integer period ratios such as 2:1, 3:2, etc.). This proximity-based threshold is commonly used in the literature when direct dynamical confirmation of resonance (via libration of resonant arguments) is not feasible due to a lack of mass constraints and orbital integration.}}
{{The threshold value of 0.02 is motivated by the work of \citet{Steffen2015}, who define near-resonant systems based on a normalized distance from exact resonance, scaled by the minimum expected variation due to resonant dynamics. Their results show that systems within $\sim$3 libration widths typically lie within $\pm$0.01--0.03 of the nominal MMR \citep[e.g. also][]{Wu2024,Batygin2017}, justifying our choice of 0.02 as a conservative, first-order approximation.}}

{{The distributions in both Fig. \ref{fig:ResoHist} and Fig. \ref{fig:CdfRratio} do not show significant differences between planet pairs near resonances and those further away, and this conclusion is supported by Kolmogorov--Smirnov (KS) tests that return no statistically significant separation between the populations.}}

Though this might be peculiar in respect to initial theoretical expectations of the conditions required for resonant capture (see the above discussion on resonant capture), various studies suggested that exoplanets may form in resonant chains which are then destabilized at later times.  With the caveat of using relatively small statistics and the question of whether this should be observed only for specific resonances, etc., this finding poses a potential challenge for our thinking of planet migration and resonant capture.  
%Planets with period ratios close to resonance
%Those systems seem to present some resonances: the counts just above and below simple integer ratios seem to be more abundant than planets with $P_{in}/P_{out}$ of a simple ratio (like 1/2, 1/8 or 1/3). This behaviour is very similar to the behaviour of asteroids in Solar System, known as Kirkwood Gaps\citep{Moons1996,Tsiganis2010} or planets with mean-motion resonances.

\subsection{Selection effect?}

%The general trend where the inner planet is smaller then the outer might be a partially result of selection effect. As figure \ref{fig:systems} indicates, "132" is more frequent configuration than "123" if the sample is based on planets above 90\% detection line. The sample is dominated by G stars (See Figure \ref{fig:stars3}), that contribute the most to the trend there "123" is more frequent than "132". This trend does not exist in M, K and F stars, as it does not hold in the de-biased sample.

The prevailing trend, where the inner planet is smaller than the outer one, may partly stem from selection biases.{{However, as shown in Figure \ref{fig:systems} (upper right corner), the "132" configuration is as frequent as the "123" configuration in the "debiased" sample, which cannot be explained by data incompleteness.
Our trends are largely influenced by G-type stars, which dominate the sample (refer to Figure \ref{fig:stars3}). M-stars are absent in the de-biased sample, and in K-planets "132" is more frequent then "123". The estimation of the influence of the data incompleteness on the derived trends can be found in section \ref{stats}.}}
%Regardless, both the "full" and "de-biased" datasets exhibit a consistent trend wherein the inner planet tends to be the smallest. This suggests that data incompleteness alone cannot account for this observed pattern.
%Notably, in M, K, and F stars, this pattern is not observed, as they are absent in the de-biased sample. 
%However, as shown in Figure \ref{fig:systems}, the "132" configuration appears more frequently than "123" in samples composed of planets above the 90\% detection threshold (de-biased). 

%In multi-planet systems, the trend is less obvious with outer compared to inner. 
%HBP: not clear what we try to say.
%The selection effect cannot explain the trend in the innermost two planets. Seem like the consequential order is relevant only to the inner two planets. The outer two out of three seem to be uniformly distributed, with almost the same probability to be in a configuration of "1,2" as "2,1". 
%Nevertheless, selection effect could not explain the trend in the innermost two planets. They do not follow a homogeneous distribution not in a full sample and nor in the de-biased one. This might indicate that the trend there inner planets tend to be smaller than  the outer ones cannot be explained by data incompleteness.

\subsection{Stellar metallicity} 
\label{sec:Metal}
\begin{figure*}
	\centering

	\includegraphics[width=1\columnwidth, trim={2.5cm 1.0cm 2.5cm 1.0cm},clip]{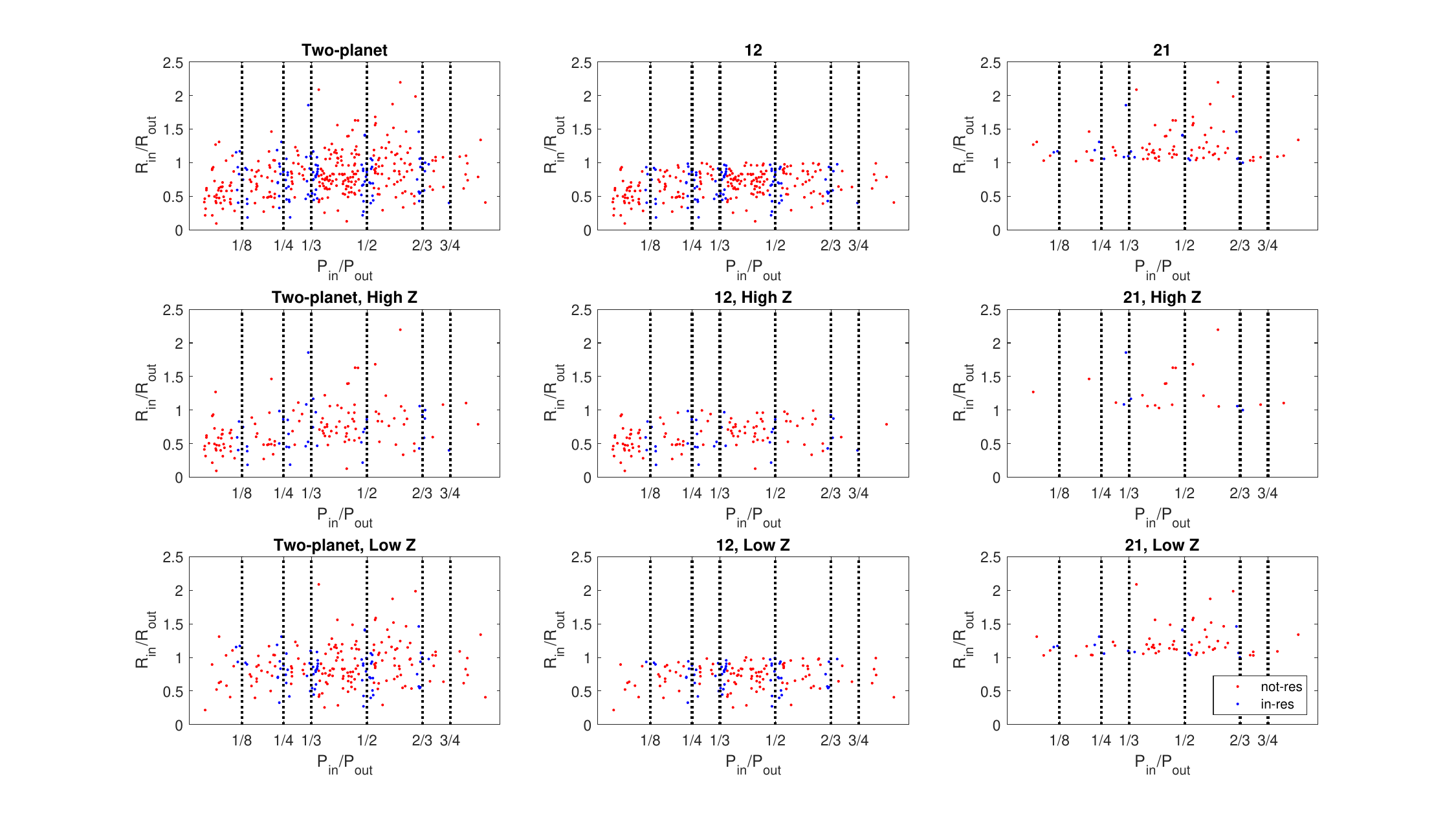}
	\caption{Ratios of planetary radii and orbital periods for two-planet systems with subdivisions. "High Z" and "Low Z" refer to planets orbiting stars with [Fe/H] above and below -0.2, respectively. Planets with period ratios close to resonances (tolerance of 0.02) are colored blue, while those not close to resonances are colored red.}
	\label{fig:ResoHistZ}
\end{figure*}

 It is reasonable to assume that the properties of a planetary system are correlated to stellar properties, or more specifically to the host metallicity, which likely relates to the metallicity of the protoplanetary disk \citep[e.g][]{Dawson2013}.  For this section, we used a "full" two-planet and three-planet sample, as if the debased sample would be divided into sub-samples, than each cannot provide sub-samples large enough for the test. We divide the sample into systems that surround high-metallicity stars and low-metallicity stars. The separation value was chosen as $[\text{Fe}/\text{H}] = \log_{10} \left( \frac{N_{\text{Fe}}}{N_{\text{H}}} \right)_{\text{star}}-\left(\frac{N_{\text{Fe}}}{N_{\text{H}}}\right)_\odot =-0.2$. This value divides the two-planet sample into two roughly equal parts, which we call "High Z" and "Low Z". The results are shown in Figures \ref{fig:CdfRratio} and \ref{fig:ResoHistZ}. The distributions of $R_{in}/R_{out}$ in Figure \ref{fig:CdfRratio} are compared quantitatively in Section \ref{subsc:KS}.
 %We compare the orbital period and planetary radius distributions of subsample.
 
 %Two-sample K-S test completely rejects the hypothesis that the period ratio of the High Z and Low Z subsamples, as well as and radii ratio of these subsamples come from the same distributions, with p values of 2.26e-04 for period ratios and 4.23e-6 for radii ratio. Therefore, the samples are not correlated and the stellar metallicity has a major impact on the distributions. 

% In this section, we used only full two-planet sample due to limitations in obtaining adequately sized sub-samples if utilizing a three-planet sample or sub-samples of two-planet systems. We divided the sample based on host stars' metallicity levels, distinguishing between those orbiting high and low metallicity stars. A separation threshold of $[\text{Fe}/\text{H}] = \log_{10} \left( \frac{N_{\text{Fe}}}{N_{\text{H}}} \right){\text{star}}-\left(\frac{N{\text{Fe}}}{N_{\text{H}}}\right)_\odot =-0.2$ was chosen, dividing the two-planet sample into roughly equal parts. The findings, presented in Figure \ref{fig:ResoHistZ}, do not indicate markedly different behaviors between the two populations.
 
%\log_{10}{\left(\frac{N_{\text{Fe}}}{N_{\text{H}}}\right)_\text{star}} - \log_{10}{\left(\frac{N_{\text{Fe}}}{N_{\text{H}}}\right)_\text{sun}}

\section{Statistical tests}
\label{stats}
{{In this section, we quantify the statistical differences and similarities of configurations of planets' populations divided by various criteria, described above. We apply different statistical tests to different aspects. }}

\subsection{\textbf{Fisher's Exact Test}}
We are interested in quantifying the statistical differences of configurations of planets' populations surrounding different stellar types (4x2 contingency table at Table \ref{tbl:2plnStars}). 
First, we compare Full and De-biased samples for each stellar type separately using  \textit{Fisher's Exact Test}, regarding "12" versus "21" counts. This test is particularly suitable here due to the small sample sizes in some categories, which makes the chi-squared test less reliable. The test was conducted for each stellar type separately, by comparing two consecutive rows from the contingency Table \ref{tbl:2plnStars} ("Full" vs. "De-biased"; Specifically, we performed the test for M-planets (rows 1 and 2), K-planets (rows 3 and 4), G-planets (rows 5 and 6), and F-planets (rows 7 and 8)). The p-values from these tests are summarized in Table \ref{tab:fisher_results}. None of the comparisons yielded a p-value below the common significance threshold of 0.05, indicating no statistically significant difference between the Full and De-biased samples for any stellar type. The p-values ranged from 0.2219 to 0.8918, with the highest observed for G-type stars and the lowest for M-type stars. These results suggest that the de-biasing process did not significantly alter the distribution of two-planet systems within any stellar type.

We examined statistical differences in two-planet system configurations around stars of different stellar types using the "Full" samples from Table \ref{tbl:2plnStars}. Fisher's Exact Test was applied to compare the distributions of "12" and "21" configurations between stellar types, with results presented in Table \ref{tab:pairwise_fisher2planet}. The null hypothesis that the samples represent distinct populations was rejected in all cases, indicating that two-planet systems are statistically similar across different stellar types.
Besides the trivial self-comparison, the highest p-value was observed when comparing F-planets to K-planets (p = 1.0), and the lowest p-value was in the comparison between F-planets and M-planets (p = 0.19). While the p-value trends might suggest some slight variations between certain stellar types, none of the differences were statistically significant, and the results are likely driven by the relatively small sample sizes in some of the stellar categories. The larger number of systems in the K-type and G-type samples may contribute to the higher p-values seen when compared to other types, whereas smaller samples (such as the M- and F-types) could lead to more variability in the test results. However, overall, the analysis shows no strong evidence for distinct configurations of two-planet systems between the different stellar types.

%Next, we are interested to test the statistical differences between configurations of comparing different stellar types, using "Full" samples. We compared all the stellar types by comparing each two stellar type. Table \ref{tab:pairwise_fisher2planet} presents the p-values from Fisher's Exact Test for each pair of stellar type. In all of the cases the null hypothesis that the sub-samples of each stellar type come from different was systematically rejected, meaning that.... (enter here). The highest p value was in comparison of F-planets to K-planets, and the lowest in comparing F-planets to M-planets,  (does it say somehign? maybe it's related to sizes?)

\begin{table}[h]
\centering
\begin{tabular}{|l|c|c|}
\hline
Stellar Type & Fisher Exact Test p-value & Reject Null Hypothesis (True/False) \\
\hline
M-planets    & 0.2219 & False  \\
K-planets    & 0.3728 & False  \\
G-planets    & 0.8918 & False  \\
F-planets    & 0.2963 & False  \\
\hline
\end{tabular}
\caption{Fisher Exact Test p-values and results for comparing Full and De-biased samples for each stellar type, from table \ref{tbl:2plnStars}. The Null Hypothesis is that two samples come from different distributions, and it was rejected systematically, meaning that the Full and De-biased samples are coming from the same distribution.} 
\label{tab:fisher_results}
\end{table}

\begin{table}[h]
\centering
\begin{tabular}{|l|c|c|c|c|}
\hline
 & M-planets Full   & K-planets Full   & G-planets Full   & F-planets Full   \\
\hline
M-planets Full   & 1.0000 (False) & 0.2909 (False) & 0.1900 (False) & 0.2803 (False) \\
K-planets Full   & 0.2909 (False) & 1.0000 (False)   & 0.5733 (False) & 1.0000 (False) \\
G-planets Full   & 0.1900 (False) & 0.5733 (False) & 1.0000 (False) & 0.7548 (False) \\
F-planets Full   & 0.2803 (False) & 1.0000 (False) & 0.7548 (False) & 1.0000 (False)  \\
\hline
\end{tabular}
\caption{Pairwise Fisher Exact Test p-values and results for Full two-planet samples across different stellar types from table \ref{tbl:2plnStars}. In brackets, there is a result of the Null Hypothesis that two samples come from different distributions: it was rejected for every pair, indicating that it is impossible to distinguish between the sub-populations.}
\label{tab:pairwise_fisher2planet}
\end{table}

We continue our analysis to examine the configurations ("12" versus "21") of two-planet systems based on absolute planetary sizes. The systems were divided into sub-samples where both planets are larger or smaller than 1.6 \(R_\oplus\), or consist of one large and one small planet. These configurations are summarized in contingency Table \ref{tbl:2pln1.6}. Fisher's Exact Test was used to compare system types, with results presented in Table \ref{tab:fisher_1.6_matrix}. Comparisons were made between full and de-biased samples of the same system type (e.g., full \(R > 1.6R_\oplus\) versus de-biased \(R > 1.6R_\oplus\)) and between different system types within the same sample category (full R $> 1.6R_\oplus$ was not compared to de-biased R $< 1.6R_\oplus$ for example). 

The test systematically failed to reject the null hypothesis (of being different) when comparing Full and De-biased samples across all cases, suggesting no significant differences between these groups for any planetary size classification. Additionally, systems with two large planets (R$>$1.6 $R_\oplus$) and two small planets (R$<$1.6 $R_\oplus$) were not found to be drawn from statistically different distributions (p=1.0). However, when comparing systems with mixed planet sizes to systems with two large planets, the null hypothesis was rejected (p=0.0048), suggesting that mixed systems (one large and one small planet) are drawn from a different distribution than systems where both planets are large. The comparison between (full) mixed systems and systems with two small planets (p=0.0523) approached the significance threshold (p=0.05), but formally failed to reject the null hypothesis. This borderline result may suggest a potential difference between these two system types, and slightly adjusting the significance threshold might yield a different conclusion.
While no significant differences were found when comparing Full vs. De-biased samples, or systems with two large vs. two small planets, the mixed-planet systems showed a significant difference when compared to large-planet systems, and a near-significant difference when compared to small-planet systems. These findings imply that mixed systems may behave differently from those with planets of uniform size, particularly when contrasted with large-planet systems. Further investigation could provide deeper insights into these distinctions.

\begin{table}[h]
\centering
\begin{tabular}{|l|c|c|c|c|c|c|}
\hline
\setlength{\tabcolsep}{0.1pt}
 & R$>$1.6$R_\oplus$ Full & R$>$1.6$_\oplus$ De-biased & R$<$1.6$_\oplus$ Full & R$<$1.6$_\oplus$ De-biased & Mixed Full & Mixed De-biased \\
\hline
R$>$1.6$R_\oplus$ Full & 1.0000 (False) & 0.6614 (False) & 1.0000 (False) & - & \textbf{0.0048 (True)} & - \\
R$>$1.6$R_\oplus$ De-biased & 0.6614 (False) & 1.0000 (False) & - & 1.0000 (False) & - & 0.4191 (False) \\
R$<$1.6$R_\oplus$ Full & 1.0000 (False) & - & 1.0000 (False) & 1.0000 (False) & \textbf{0.0523 (False?)} & - \\
R$<$1.6$R_\oplus$ De-biased & - & 1.0000 (False) & 1.0000 (False) & 1.0000 (False) & - & 1.0000 (False) \\
Mixed Full & \textbf{0.0048 (True)} & - & \textbf{0.0523 (False?)} & - & 1.0000 (False) & - \\
Mixed De-biased & - & 0.4191 (False) & - & 1.0000 (False) & - & 1.0000 (False) \\
\hline
\end{tabular}
\caption{Fisher Exact Test p-values and results for planetary systems divided by planet size classifications, comparing "12" versus "21" of Full and De-biased samples, from Table \ref{tbl:2pln1.6}. Full and De-biased of different types were not compared. Bold indicates the cases where the Null Hypothesis of no difference in the distribution was rejected, indicating a different origin of the populations. Then, comparing Mixed Full with $R<1.6R_\oplus$, the result was very close to rejection of the Null Hypothesis.}
\label{tab:fisher_1.6_matrix}
\end{table}

\subsection{\textbf{Anderson–Darling test} }
\label{subsc:KS}
%\Hagai{We should repeat the same type of test method for the 3 planet ordering, while sampling from the sub two-planets distribution to see whether the 3-planet ordering has additional information beyond that given by the 2-planet orderings}

%Next we compare the distribution of planetary radii ratios, $R_{in}/R_{out}$, there $R_{in}$ is the radius of an inner planet out of the given pair and $R_{out}$ is the radius of the outer. As can be seen from Figure \ref{fig:Subsystems}, in each system the larger planets tend to be further from their host stars compared to smaller ones. This hypothesis can be examined using a two-sample Kolmogorov–Smirnov test (K–S test). The two-sample test (not to be confused with one-sample K-S test) compare distributions. The two-sample K-S test compares the distributions of two samples. Each test assumes a null-hypothesis that two tested samples originate from the same distribution. We use it to compare distributions of a ratios between the radii of the inner planets and that of the outer ones $R_{in}/R_{out}$ of different samples.
Next, we compare the distribution of planetary radii ratios, $R_{in}/R_{out}$, where $R_{in}$ is the radius of an inner planet in a given pair and $R_{out}$ is the radius of the outer planet. As seen in Figure \ref{fig:Subsystems}, larger planets tend to be further from their host stars compared to smaller ones. This hypothesis can be tested quantitatively.

%This hypothesis can be examined using a two-sample Kolmogorov–Smirnov (K–S) test, which compares the distributions of two samples(not to be confused with a one-sample K-S test). Each test assumes a null hypothesis that the two samples originate from the same distribution \footnote{The output of the two-sample Kolmogorov–Smirnov includes the \textit{p-value}. The \textit{p-value} indicates the probability of obtaining a test statistic at least as extreme as the one observed, assuming that the null hypothesis is true. A low p-value suggests strong evidence against the null hypothesis.}. We use it to compare the distributions of the ratios between the radii of the inner and outer planets, $R_{in}/R_{out}$, across different samples.
We perform the following procedure:

To start, we create a synthetic "homogeneous" distribution based on two-planet systems. This distribution will help us test a simple null hypothesis, which suggests that the distribution of radius ratios is uniform. In other words, a hypothesis that assumes that the inner planet is just as likely to have a larger radius than the outer planet as the outer planet is to have a larger radius than the inner planet. This synthetic sample is constructed by randomly shuffling the periods and radii within each pair of the two-planet samples. This process is repeated 100 times, resulting in a total of 34,600 pairs of (\textit{P},\textit{R}), while preserving the original marginal distribution of planetary periods and radii.

Next, we calculate the ratio of the inner planet to the outer planet, $R_{in}/R_{out}$, for observed pairs in the two-planet systems (both full and de-biased), for pairs in three-planet systems (both full and de-biased), and for the synthetic homogeneous sample described here. Afterwards, we compare the distributions of $R_{in}/R_{out}$ between each two samples, using a two-sample or K-sample Anderson–Darling test\footnote{The K-sample Anderson–Darling test compares whether multiple samples come from the same distribution. The null hypothesis is that the samples are drawn from the same distribution. If the p-value is low (below 0.05), the null hypothesis is rejected, indicating that the samples are significantly different. If the p-value is high (above 0.05), the test fails to reject the null hypothesis, indicating that the samples are likely from the same distribution.}
The comparisons are repeated between every pair of sub-samples, with the results shown in Table \ref{tab:KS}. The p-values are listed, and bold values indicate failure to reject the null hypothesis at the 5$\%$ significance level, meaning the two distributions are indistinguishable. 

We assessed each sample against the synthetic "homogeneous" distribution to determine if it came from that distribution. This hypothesis was consistently disproved for every sample, consistent with the ordering being non-random (see also \cite{Kipping2017}).{{The highest similarity (i.e., failure to reject the null hypothesis; indicated by higher p-values) is observed between full systems and their de-biased counterparts within the same configuration, as shown by the bold p-values. This outcome is expected, as de-biased systems represent a close subset of the full systems. In addition, the test did not find a signification difference between $R_{in}/R_{out}$ distributions of ab/3 and ac/3 both full and debased), while it found a sharp distinction to bc/3. This means that the last two planets' ratio differs significantly from the others. In addition, Two-planet systems are indistinguishable from ab/3, indicating that the first two planets out of three come from a very similar distribution as two-planet systems, and suggesting that two-planet systems might have a third companion to be discovered. }}

\begin{table}[]
\centering
\setlength{\tabcolsep}{2pt} % Reduced column spacing
\begin{tabular}{|c|c|ccccccc|}
\hline
                    & Homogeneous & \multicolumn{1}{c|}{Two-planet (full)} & \multicolumn{1}{c|}{Two-planet (debias)} & \multicolumn{1}{c|}{ab/3 (full)} & \multicolumn{1}{c|}{ab/3 (debias)} & \multicolumn{1}{c|}{ac/3 (full)} & \multicolumn{1}{c|}{ac/3 (debias)} & bc/3 (full) \\ \hline
Two-planet (full)   & \textbf{0.001}       & \textbf{}                      & \textbf{}                      & \textbf{}                      & \textbf{}                      &                               &                               &          \\
                    & (Diff.)      &                               &                               &                               &                               &                               &                               &          \\ \cline{1-3}
Two-planet (debias) & \textbf{0.001}       & \multicolumn{1}{c|}{$>$0.25} & \textbf{}                      & \textbf{}                      & \textbf{}                      &                               &                               &          \\
                    & (Diff.)      & \multicolumn{1}{c|}{(Same)}   &                               &                               &                               &                               &                               &          \\ \cline{1-4}
ab/3 (full)         & \textbf{0.001}       & \multicolumn{1}{c|}{$>$0.25} & \multicolumn{1}{c|}{$>$0.25} & \textbf{}                      & \textbf{}                      & \textbf{}                      & \textbf{}                      &          \\
                    & (Diff.)      & \multicolumn{1}{c|}{(Same)}   & \multicolumn{1}{c|}{(Same)}   &                               &                               &                               &                               &          \\ \cline{1-5}
ab/3 (debias)       & \textbf{0.001}       & \multicolumn{1}{c|}{$>$0.25} & \multicolumn{1}{c|}{$>$0.25} & \multicolumn{1}{c|}{$>$0.25} & \textbf{}                      & \textbf{}                      & \textbf{}                      &          \\
                    & (Diff.)      & \multicolumn{1}{c|}{(Same)}   & \multicolumn{1}{c|}{(Same)}   & \multicolumn{1}{c|}{(Same)}   &                               &                               &                               &          \\ \cline{1-6}
ac/3 (full)         & \textbf{0.001}       & \multicolumn{1}{c|}{\textbf{0.017}}    & \multicolumn{1}{c|}{\textbf{0.017}}    & \multicolumn{1}{c|}{0.203}           & \multicolumn{1}{c|}{0.203} & \textbf{}                      & \textbf{}                      &          \\
                    & (Diff.)      & \multicolumn{1}{c|}{(Diff.)}   & \multicolumn{1}{c|}{(Diff.)}   & \multicolumn{1}{c|}{(Same)}   & \multicolumn{1}{c|}{(Same)}   &                               &                               &          \\ \cline{1-7}
ac/3 (debias)       & \textbf{0.001}       & \multicolumn{1}{c|}{\textbf{0.017}}    & \multicolumn{1}{c|}{\textbf{0.017}}    & \multicolumn{1}{c|}{0.203}           & \multicolumn{1}{c|}{0.203} & \multicolumn{1}{c|}{$>$0.25} & \textbf{}                      &          \\
                    & (Diff.)      & \multicolumn{1}{c|}{(Diff.)}   & \multicolumn{1}{c|}{(Diff.)}   & \multicolumn{1}{c|}{(Same)}   & \multicolumn{1}{c|}{(Same)}   & \multicolumn{1}{c|}{(Same)}   &                               &          \\ \cline{1-8}
bc/3 (full)         & \textbf{0.004}       & \multicolumn{1}{c|}{\textbf{0.001}}    & \multicolumn{1}{c|}{\textbf{0.001}}    & \multicolumn{1}{c|}{\textbf{0.001}}    & \multicolumn{1}{c|}{\textbf{0.001}}    & \multicolumn{1}{c|}{\textbf{0.001}}    & \multicolumn{1}{c|}{\textbf{0.001}}    & \textbf{}          \\
                    & (Diff.)      & \multicolumn{1}{c|}{(Diff.)}   & \multicolumn{1}{c|}{(Diff.)}   & \multicolumn{1}{c|}{(Diff.)}   & \multicolumn{1}{c|}{(Diff.)}   & \multicolumn{1}{c|}{(Diff.)}   & \multicolumn{1}{c|}{(Diff.)}   &          \\ \hline
bc/3 (debias)       & \textbf{0.004}       & \multicolumn{1}{c|}{\textbf{0.001}}    & \multicolumn{1}{c|}{\textbf{0.001}}    & \multicolumn{1}{c|}{\textbf{0.001}}    & \multicolumn{1}{c|}{\textbf{0.001}}    & \multicolumn{1}{c|}{\textbf{0.001}}    & \multicolumn{1}{c|}{\textbf{0.001}}    & $>$0.25 \\
                    & (Diff.)      & \multicolumn{1}{c|}{(Diff.)}   & \multicolumn{1}{c|}{(Diff.)}   & \multicolumn{1}{c|}{(Diff.)}   & \multicolumn{1}{c|}{(Diff.)}   & \multicolumn{1}{c|}{(Diff.)}   & \multicolumn{1}{c|}{(Diff.)}   & (Same)   \\ \hline
\end{tabular}

\caption{Anderson-Darling test p-values for pairwise comparisons of the samples' Radii ratios ($R_{in}/R_{out}$) distributions. Bold values indicate a rejection the null hypothesis that the two samples are drawn from the same distribution, implying that the two compared distributions are statistically distinguishable. (Diff.) = statistically significant difference. (Same)= the test did not indicate a significant difference.}
\label{tab:KS}
\end{table}

{{Table \ref{tab:AD_metal} shows p-values of Anderson-Darling test for pairwise comparisons of the samples' Radii ratios ($R_{in}/R_{out}$) distributions for Two-planet sample and its sub-samples, taking into account Stellar metallicity or no. It shows that $R_{in}/R_{out}$ distribution of "21" configuration is the same as the full one, and High-Z and Low-Z of "21" follow the same distribution. However, this might be a result of relatively low statistics.}}

\begin{table}[]
\centering
\setlength{\tabcolsep}{2pt}
\begin{tabular}{c|c|ccccccc|}
\cline{2-9}
                                         & Two-planet     & \multicolumn{1}{c|}{Two-planet, High Z} & \multicolumn{1}{c|}{Two-planet, Low Z} & \multicolumn{1}{c|}{12}             & \multicolumn{1}{c|}{12, High Z}     & \multicolumn{1}{c|}{12, Low Z}      & \multicolumn{1}{c|}{21}      & 21, High Z \\ \hline
\multicolumn{1}{|c|}{Two-planet, High Z} & \textbf{0.001} &                                         &                                        &                                     &                                     &                                     &                              &            \\
\multicolumn{1}{|c|}{}                   & (Diff.)         &                                         &                                        &                                     &                                     &                                     &                              &            \\ \cline{1-3}
\multicolumn{1}{|c|}{Two-planet, Low Z}  & \textbf{0.025} & \multicolumn{1}{c|}{\textbf{0.001}}     &                                        &                                     &                                     &                                     &                              &            \\
\multicolumn{1}{|c|}{}                   & (Diff.)         & \multicolumn{1}{c|}{(Diff.)}             &                                        &                                     &                                     &                                     &                              &            \\ \cline{1-4}
\multicolumn{1}{|c|}{12}                 & \textbf{0.001} & \multicolumn{1}{c|}{\textbf{0.002}}     & \multicolumn{1}{c|}{\textbf{0.001}}    &                                     &                                     &                                     &                              &            \\
\multicolumn{1}{|c|}{}                   & (Diff.)         & \multicolumn{1}{c|}{(Diff.)}             & \multicolumn{1}{c|}{(Diff.)}            &                                     &                                     &                                     &                              &            \\ \cline{1-5}
\multicolumn{1}{|c|}{12, High Z}         & \textbf{0.001} & \multicolumn{1}{c|}{\textbf{0.005}}     & \multicolumn{1}{c|}{\textbf{0.001}}    & \multicolumn{1}{c|}{\textbf{0.010}} &                                     &                                     &                              &            \\
\multicolumn{1}{|c|}{}                   & (Diff.)         & \multicolumn{1}{c|}{(Diff.)}             & \multicolumn{1}{c|}{(Diff.)}            & \multicolumn{1}{c|}{(Diff.)}         &                                     &                                     &                              &            \\ \cline{1-6}
\multicolumn{1}{|c|}{12, Low Z}          & \textbf{0.001} & \multicolumn{1}{c|}{\textbf{0.001}}     & \multicolumn{1}{c|}{\textbf{0.001}}    & \multicolumn{1}{c|}{\textbf{0.023}} & \multicolumn{1}{c|}{\textbf{0.001}} &                                     &                              &            \\
\multicolumn{1}{|c|}{}                   & (Diff.)         & \multicolumn{1}{c|}{(Diff.)}             & \multicolumn{1}{c|}{(Diff.)}            & \multicolumn{1}{c|}{(Diff.)}         & \multicolumn{1}{c|}{(Diff.)}         &                                     &                              &            \\ \cline{1-7}
\multicolumn{1}{|c|}{21}                 & \textbf{0.001} & \multicolumn{1}{c|}{\textbf{0.001}}     & \multicolumn{1}{c|}{\textbf{0.001}}    & \multicolumn{1}{c|}{\textbf{0.001}} & \multicolumn{1}{c|}{\textbf{0.001}} & \multicolumn{1}{c|}{\textbf{0.001}} &                              &            \\
\multicolumn{1}{|c|}{}                   & (Diff.)         & \multicolumn{1}{c|}{(Diff.)}             & \multicolumn{1}{c|}{(Diff.)}            & \multicolumn{1}{c|}{(Diff.)}         & \multicolumn{1}{c|}{(Diff.)}         & \multicolumn{1}{c|}{(Diff.)}         &                              &            \\ \cline{1-8}
\multicolumn{1}{|c|}{21, High Z}         & \textbf{0.001} & \multicolumn{1}{c|}{\textbf{0.001}}     & \multicolumn{1}{c|}{\textbf{0.001}}    & \multicolumn{1}{c|}{\textbf{0.001}} & \multicolumn{1}{c|}{\textbf{0.001}} & \multicolumn{1}{c|}{\textbf{0.001}} & \multicolumn{1}{c|}{$>$0.25} &            \\
\multicolumn{1}{|c|}{}                   & (Diff.)         & \multicolumn{1}{c|}{(Diff.)}             & \multicolumn{1}{c|}{(Diff.)}            & \multicolumn{1}{c|}{(Diff.)}         & \multicolumn{1}{c|}{(Diff.)}         & \multicolumn{1}{c|}{(Diff.)}         & \multicolumn{1}{c|}{(Same)}  &            \\ \hline
\multicolumn{1}{|c|}{21, Low Z}          & \textbf{0.001} & \multicolumn{1}{c|}{\textbf{0.001}}     & \multicolumn{1}{c|}{\textbf{0.001}}    & \multicolumn{1}{c|}{\textbf{0.001}} & \multicolumn{1}{c|}{\textbf{0.001}} & \multicolumn{1}{c|}{\textbf{0.001}} & \multicolumn{1}{c|}{$>$0.25} & $>$0.25    \\
\multicolumn{1}{|c|}{}                   & (Diff.)         & \multicolumn{1}{c|}{(Diff.)}             & \multicolumn{1}{c|}{(Diff.)}            & \multicolumn{1}{c|}{(Diff.)}         & \multicolumn{1}{c|}{(Diff.)}         & \multicolumn{1}{c|}{(Diff.)}         & \multicolumn{1}{c|}{(Same)}  & (Same)     \\ \hline
\end{tabular}

\caption{Same as Table~\ref{tab:KS}, comparing sub-samples of Two planet systems (full). "High Z" and "Low Z" refer to stellar metallicity above and below $-0.2$ (see section \ref{sec:Metal} and Figure \ref{fig:CdfRratio}).}
\label{tab:AD_metal}
\end{table}

\section{Discussion}

Our analysis of \textit{Kepler} multi-planet systems reveals a strong tendency for inner planets to be smaller than outer planets within a given system. This trend persists in both the "full" and "de-biased" samples, suggesting that it is not solely a product of observational biases. The size ordering is more pronounced in inner planet pairs compared to outer pairs, particularly in three-planet systems. In addition, we find larger inner-to-outer planet size ratios for systems with larger period ratios, but we do not find a strong dependence of the size ratios for planet pairs in resonances (though the statistics might be too small). Finally, we find a strong metallicity dependence of both size ratios and period ratios of planetary pairs. 

These findings have important implications for our understanding of planet formation and evolution processes, but their exact interpretation requires comparison with theoretical studies about planetary ordering, which have yet to be done. 

\subsection{Influence of Stellar Properties and metallicity}
Our analysis of two-planet systems suggests a potential correlation between planetary size ordering and stellar metallicity. The statistical test results indicate that the period ratios and radii ratios of planets orbiting high-metallicity stars differ significantly from those orbiting low-metallicity stars (see table \ref{tab:AD_metal} and figure \ref{fig:CdfRratio}). This finding supports the idea that the initial conditions of planet formation, influenced by the composition of the protoplanetary disk, play a role in determining the final architecture of planets within a system. As mentioned above, one might expect lower metallicity environments to be quieter and better represent initial conditions, on average, compared with later, more dynamical evolution due to scattering that might be more preferred in higher metallicity environments. Nevertheless, given the required alignment of the planet orbits in the Kepler sample of multi-planet systems, the whole sample might arise from quite evolved systems allowing for less-excited inclinations, as we further discuss below.

\subsection{Observational Biases}

Despite our efforts to account for observational biases through the use of \textit{Kepler}'s detection limits \citep{Petigura2013}, it is important to acknowledge the limitations of our current dataset. The inner system bias inherent in transit surveys means that our understanding of size ordering in the outer regions of planetary systems remains incomplete \citep{Winn2015}. Future studies should aim to incorporate data from other detection methods, such as radial velocity surveys and microlensing, to provide a more comprehensive view of planetary system architectures.

Additionally, the potential influence of multiplicity bias \citep{Zhu2018} and the known correlation between planet mass and orbital period \citep{Bashi2017} should be carefully considered in future analyses. Long-term surveys focused on detecting outer planets will be crucial for completing our picture of system architectures and testing the universality of the trends observed in this study.

Finally, systems with more detectable planets are more likely to be nearly coplanar, which may not be representative of the general population of planetary systems \citep{Zhu2018}, and in particular, might suggest more "quiet" systems in which planet-planet scattering might not have played a major role. If this is the case, such systems better represent the initial conditions, but might not show the effects of other processes that might be frequent in other systems. Similar analysis of RV-detected systems and possible differences that they might show in comparison with the transiting planets samples could reveal the unique role played by these other processes, and distinguish selection biases from actual planetary formation processes.  

\subsection{\textbf{Comparison to one-planet systems}}

 \begin{figure*}
	\centering

	\includegraphics[width=0.5\columnwidth, trim={3.0cm 8.2cm 4.0cm 8.3cm},clip]{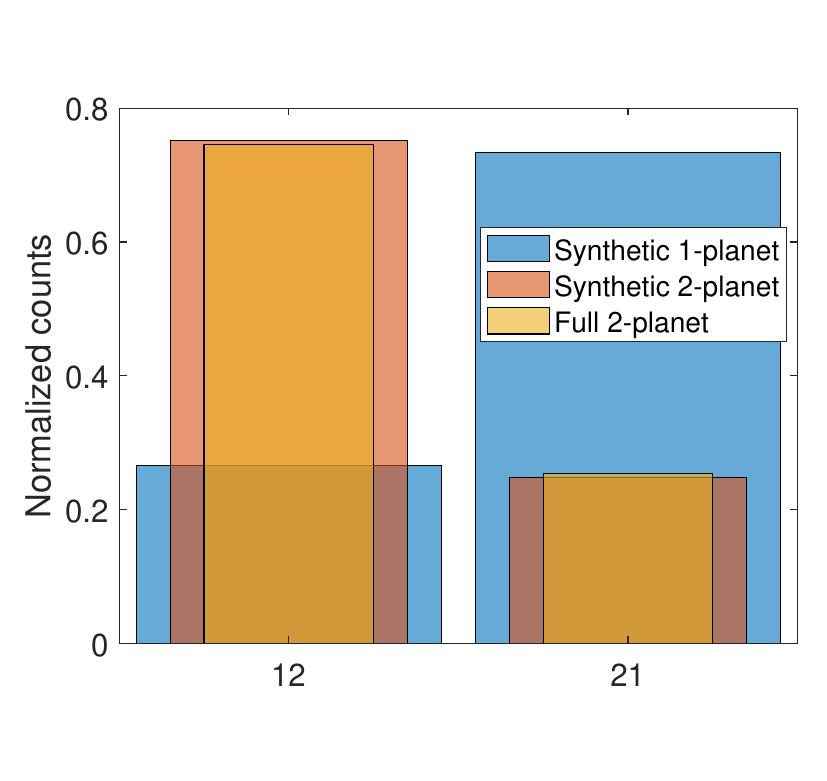}
	\caption{ Normalized distributions of synthetic two-planet systems made of one-planet systems ("Synthetic 1-planet") and two-planet systems ("Synthetic 2-planet") and real  "full" two-planet sample. }
	\label{fig:OneTwoPlanetRAND}
\end{figure*}

{{To test the influence of observational biases on the ordering of planets in two-planet systems, we conducted a comparison between observed systems and synthetic systems generated through randomized sampling. For this section, a sample of synthetic two-planet systems were constructed by pairing planets drawn from systems with a single detected planet. During the random selection process, we ensured that planets were hosted by stars of similar mass (with differences constrained to within $2\%$ by mass), and that the pairs satisfied Gladman’s stability criterion, which requires the separation between the planetary semi-major axes to exceed 3.5 times the mutual Hill radius. This sample is referred to as "Synthetic 1-planet".}}

\begin{table}[h]
\centering
\begin{tabular}{|c|c|c|c|}
\hline
 & 12 & 21 & Total \\
\hline
Synthetic 1-planet & 266 (26.60\%) & 734 (73.40\%) & 1000 \\
Synthetic 2-planet & 751 (75.10\%) & 249 (24.90\%) & 1000 \\
Full 2-planet & 273 (74.59\%) & 93 (25.41\%) & 366 \\
\hline
\end{tabular}
\caption{Contingency table for two-planet systems comparing Synthetic 1-planet, Synthetic 2-planet, and Full 2-planet.}
\label{tbl:synt2pl}
\end{table}

{{In a similar way, another sample of synthetic systems was built, by randomly selecting planets from a two-planet sample and assigning them to be new pairs. This sample is referred to as "Synthetic 2-planet". For each synthetic sample, we compare the configurations of "12" versus "21". The results are presented in Figure \ref{fig:OneTwoPlanetRAND} and contingency table \ref{tbl:synt2pl}.}}

{{A direct comparison of the period and radius ratios between the observed and synthetic samples revealed statistically significant differences in the distributions, showing that single-planet systems do not come from the same distribution as two-planet systems. This finding indicates that the ordering of planets in observed two-planet systems cannot be fully attributed to random placement or observational biases, as the synthetic sample was constructed to replicate these biases. The results suggest that the observed configurations may arise from intrinsic physical or dynamical processes during planetary formation or evolution.}}

\subsection{Implications for Planet Formation Models}

Our findings provide important constraints for planet formation models. The strong tendency for size ordering, particularly in the inner regions of planetary systems, suggests that models must account for both the initial formation locations of planets and their subsequent migration histories. The variations observed in outer planet pairs highlight the need for models to incorporate the complex dynamical interactions that can occur during and after the formation phase.

Future theoretical work should focus on developing more sophisticated models that can simultaneously explain the observed size orderings, the formation of compact systems, and the diversity of architectures seen in the exoplanet population. Such models will need to consider the interplay between core accretion, disk migration, planet-planet scattering, and long-term dynamical evolution to fully capture the processes shaping planetary system architectures.

In conclusion, our study of planet size ordering in Kepler multi-planet systems provides valuable insights into the processes that shape planetary system architectures. The observed trends support models of planet formation and migration while highlighting the complex interplay of various dynamical processes. As our understanding of these systems continues to evolve, future observations and theoretical work will be crucial in unraveling the full story of how planetary systems form and evolve.

\section{Summary}
\label{sec:Conclusions}
This study analyzed the ordering of planet sizes within multi-planetary systems, focusing on systems with two to three planets, with some consideration given to four-planet systems. Using data from the NASA Exoplanet Archive and addressing selection biases identified by \citet{Petigura2013}, we find non-random and non-trivial size ordering of planetary system, which also depends on the system metallicity and planet-multiplicity, and overall planet sizes. In particular, planet-ordering can be a key novel and unique observable to study and constrain planetary formation and evolution models of exoplanet systems and should be considered in any comparison of their observations. Furthermore, it calls for planet formation theories to provide predictions for this observable, which, to the best of our knowledge, have not been done yet.

Our main results can be summarized in the following.
\begin{enumerate}
    \item \textbf{Tendency for Inner Planets to be Smaller}:
    The analysis reveals a clear trend: in a typical system, the inner planets are generally smaller than the outer planets. This observation holds for both the full sample and the de-biased sample, suggesting that this is not merely a result of observational bias but reflects an intrinsic property of planetary systems. This finding has important implications for our understanding of planet formation and evolution, as size ordering aligns with models of planetary formation and migration, where larger planets form farther out in the protoplanetary disk and migrate inward over time.

    \item \textbf{Variation Across Different System Multiplicities}:
    For systems with three planets, the tendency for size ordering (inner planets being smaller) is also observed, but with more variation. This variation suggests that additional dynamical processes may influence the size distribution in more complex systems. Further research into these systems could provide deeper insights into the mechanisms at play.

    \item \textbf{Dependence on stellar and planetary types}:
    Our results suggest differences exist between the ordering of planetary systems with different host types {{(figures \ref{fig:stars2}-\ref{fig:stars3})}}, as well as differences between the ordering of gaseous and terrestrial planets {{(figures \ref{fig:planet2}-\ref{fig:planet3})}}. It is still difficult to say whether some of these arise from small statistics and observational biases that might still exist, but these are important to note and further explore with larger databases.
    
    \item \textbf{Impact of Data Completeness}:
    By comparing the full sample with the de-biased sample, it is evident that data incompleteness affects the detection of smaller planets, and although the overall trend of inner planets being smaller remains significant in two-planet systems, considering a debiased sample significantly affects the results for three-planet systems and in particular the outer planetary pair ordering. This finding underscores the importance of correcting detection biases in exoplanet studies to uncover genuine astrophysical trends.
    
    \item \textbf{Potential Influences of Stellar Metallicity}:
    The study shows a dependence of the planetary ordering and architecture on metallicity, suggesting that planets around metal-rich stars may show different size hierarchies compared to those around metal-poor stars. This aspect warrants further investigation, as it could link the initial conditions of planet formation to the observed architectures of planetary systems.

    \item \textbf{Dependence of ordering on period ratios}: We find the size-ratio distribution depends on the period ratios non-trivially. In particular, we see a trend for planetary systems with smaller inner-to-outer size ratios to prefer smaller period ratios. However, the distribution of the size ratio of planet pairs in resonance does not show a significant difference between planet pairs far from resonance. This might be in tension with the expectation that resonance capture requires more fine-tuned conditions and different planet mass ratios (e.g., requiring a more massive outer planet for convergent migration to take place and lead to resonant capture). This calls for further study, but in any case highly unique opportunities for exploring planetary dynamical processes using the novel observable of planetary ordering. 
    
\end{enumerate}

The findings have significant implications for understanding planet formation and evolution. 
{{Our}} results can refine existing theories and guide future observational campaigns to target specific configurations predicted by theoretical models.{{The consistent trend of smaller inner planets supports models of atmospheric photoevaporation, where high-energy stellar radiation strips away the gaseous envelopes of close-in planets, preferentially reducing their radii over time}} \citep[e.g.][]{Owen2013,Lopez2013,Jin2014}.

%The consistent trend of smaller inner planets supports models where planetary migration and in-situ formation play crucial roles. 

Future studies on the topic should aim to expand the dataset to include more multi-planet systems with well-characterized biases, to investigate the role of stellar properties, such as metallicity and age, in influencing planetary ordering, and possibly to utilize statistical methods to disentangle the effects of formation and dynamical evolution processes. Furthermore, given the differences in detection biases, a similar study of multi-planet systems detected by radial-velocity measurement would provide an important and complementary picture.

\section{Acknowledgements}
HBP would like to acknowledge support from the Minerva Center for Life under extreme planetary conditions. We thank the reviewers for their constructive feedback and insightful suggestions, which greatly improved the quality and clarity of this work. 

%\appendix
%One might ask why similar methodology can be applied to our sample using hierarchy of planetary masses. This is shown in Figure \ref{fig:systemsMass}. In this case the amount of detected planets does not allow to perform a full statistical study. However, it seems like the tendancy is toward inner planet to smaller in mass exists. This was not chacked against selection effects, as the masses comes from mesumrents by different instruments. 

%Considering whether a similar approach can be applied to our sample's masses instead of radii:  hierarchy of planetary masses as depicted in Figure \ref{fig:systemsMass}. Unfortunately, the limited number of detected planets prevents a thorough statistical analysis. Nonetheless, a trend seems apparent, indicating a preference for smaller masses among inner planets. However, the assertion remains unverified against potential selection biases, given the diverse array of instruments utilized for mass measurements.

%\begin{figure}[h]
%	\centering

%	\includegraphics[width=0.5\columnwidth, trim={0.4cm 6.9cm 0.4cm 7.0cm},clip]{CompareSystemsMass.pdf}
	%\caption{Histogram of distribution of planetary   relative masses for two-, three- and four-planet systems.}
	%\label{fig:systemsMass}
%\end{figure}

%\newpage
\bibliography{references}
\bibliographystyle{aasjournal}

\end{document}